\documentclass{aastex}
\usepackage{emulateapj5}

\newcommand\rosat{{\it ROSAT }}

\newcommand\ttwo{$\times10^{-2}$}
\newcommand\tthr{$\times10^{-3}$}

\newcommand\wf{wavefunction }
\newcommand\tb{{tightly-bound }}
\newcommand\lb{{loosely-bound }}

\begin{document}
\submitted{Accepted for publication in the Astrophysical Journal}
\title{Atomic calculation for the atmospheres of strongly-magnetized
neutron stars}
\author{Kaya Mori and Charles J. Hailey}
\affil{Columbia Astrophysics Laboratory, New York, NY 10027}
\begin{abstract}
Complete  modeling of  radiative transfer  in neutron  star atmospheres  is in
progress,  taking into  account  the anisotropy  induced  by magnetic  fields,
non-ideal effects and general relativity.  As part of our modeling, we present
a  novel atomic  calculation method  producing  an extensive  atomic data  set
including  energy values  and  oscillator strengths  in  the so-called  Landau
regime ($B  > 4.7\times10^9Z^2$ G).  Conventional atmosphere models  for $B=0$
are not  applicable to  typical field strengths  of cooling neutron  stars ($B
\sim10^{12}-10^{13}$  G),  since  an   atom  no  longer  keeps  its  spherical
shape. The elemental  composition and the configuration of  the magnetic field
in the atmosphere are presently unknown,  so that atomic data must be produced
for  ground and  excited states  of  several ions  as a  function of  magnetic
field.  To accomplish  this efficiently,  we minimized  the iterations  in the
Hartree  equation and  treated  exchange  terms and  higher  Landau states  by
perturbation methods. This  method has the effect of  reducing the computation
time  significantly. Inclusion  of higher  Landau  states gives  us much  more
accurate data for  inner orbitals unlike other methods  based on the adiabatic
approximation. While existing  atomic data in the Landau  regime are available
only for  low $Z$  atoms, our method  can be  used in elements  up to  Fe with
sufficient  accuracy to  be of  use for  spectroscopic missions  such  as {\it
Chandra}, {\it XMM-Newton} and next-generation X-ray telescopes.
\end{abstract}
\slugcomment{Accepted for publication in the Astrophysical Journal}

% ---------------------------------------------------------------------------

\section{Introduction}
\label{intro}

Since  their discovery in  1967 \citep{hewish68},  pulsars have  attracted the
interest of researchers  in various fields due to  their exotic properties and
extreme physical conditions.  Although many of the pulsars  were discovered in
the  radio  band, multi-wavelength  studies  have  helped  to elucidate  their
properties. \rosat discovered thermal components ($T\sim10^5-10^6$ K) from the
so-called    cooling    neutron    stars    in    the    soft    X-ray    band
\citep{becker97}. Emergent spectra from the photosphere can be modified from a
pure blackbody spectrum by the presence of an atmosphere \citep{romani87}. The
geometrical thickness of neutron star atmospheres is small ($\sim 0.1-100$ cm)
due to the strong gravitational field \citep{pavlov95_1}. Nevertheless, it can
be optically thick for reprocessing thermal photons due to the high density.

With great improvements in the  sensitivity and the resolution of recent X-ray
telescopes, X-ray spectroscopy can directly provide information on the surface
of neutron stars. There are  two possible implications that the atmosphere may
have on the X-ray spectrum.  Precise measurement of surface temperature taking
into account  the atmospheric effects  is crucial for determining  the cooling
curves  of  neutron  stars,   which  constrains  various  equations  of  state
\citep{tsuruta95,  page98,  yakovlev99}.   Secondly,  atomic  (line  or  edge)
absorption  features  along  with   cyclotron  lines  enable  us  to  diagnose
properties  of the  atmosphere.

Magnetic field ($B$) can be  measured directly by the observation of cyclotron
lines \citep{trumper, wheaton}, or indirectly  by the spindown rate of pulsars
assuming  a  dipole magnetic  configuration  \citep{ostriker69}.  Most of  the
observed pulsars are  believed to possess very strong  magnetic fields ($B\sim
10^{11}-10^{13}$ G), except the  millisecond pulsars whose magnetic fields are
relatively  low ($B\sim10^{8}-10^{10}$  G) \citep{camilo94}.  Surface elements
($Z$) are not  known at present. Strong gravity  stratifies the surface layers
so  that  the   lightest  element  resides  at  the   top  of  the  atmosphere
\citep{alcock80}. However, any element between H and Fe is feasible. Accretion
from the  interstellar medium could accumulate  a hydrogen layer,  while an Fe
atmosphere is favored  from the study of the  mass-cut in supernova explosions
\citep{woosley86}.  Lighter elements  are formed  by the  spallation  of heavy
nuclei by particles \citep{miller90}. On  the other hand,
pycnonuclear  reactions   could  produce  heavier  nuclei   from  light  nuclei
\citep{lai97,  salpeter98}. The fallback  from the  outer shells  of supernova
remnants  could be  of any  element \citep{chevalier96}.  The  surface density
($\rho$) is also  uncertain in the range of  $\rho\sim 0.1-10\mbox{ gcm}^{-3}$
\citep{pavlov95_1}. The state of matter  on the surface could be liquid, solid
or  in the form  of molecular  chains along  the field  \citep{abrahams, lai92,
demeur94,  ortiz95,  lai96, lai97}.  In high  density plasma, atoms are strongly-coupled, inducing non-ideal  effects, and  electrons may  be partially degenerate.

% =============================================================================

\section{Atmosphere models}
\label{atmos_model}

There  have  been  several   attempts  at  modeling  magnetized  neutron  star
atmospheres. For weakly-magnetized pulsars  ($B\le 10^9$ G), atmosphere models
for  $B=0$  are  applicable  \citep{romani87,  rajagopal96,  zavlin96}.  These
non-magnetic atmosphere models  have been applied to several  neutron stars 
\citep{zavlin98,  pons00}. However,  for  the strongly-magnetized  atmospheres
($B>10^9$  G), atomic models  require a  new scheme  quite different  from the
conventional  atomic  calculations  for  the  $B=0$ case.  For  instance,  the
ionization  threshold of a hydrogen atom  increases  to 160  eV at  $B=10^{12}$
G. Also,  in the  presence of  a strong magnetic  field photon  opacities are
highly dependent  on the direction of the magnetic field and  the polarization of
photons  \citep{pavlov95_1}. Thermodynamic  properties of  hydrogen
atmospheres have been  intensively investigated by  \citet{lai95}, \citet{steinberg98} and
\citet{potekhin99}.  They  have  computed  the equation  of state  and  the degree  of
ionization assuming LTE and  taking into account coupling effects such
as the motional Stark field and the pressure ionization peculiar to a
strongly-magnetized dense plasma. The hydrogen atmosphere model
of  \citet{potekhin99}  is  constructed  based  on the accurate  atomic  database
obtained  by \citet{potekhin98-2}.  Recently, \citet{ozel01}  and \citet{ho01}
studied spectral  signatures from  atmospheres of strongly  magnetized neutron
stars, utilizing improvements in radiative transfer theory.  However, they
assumed fully-ionized hydrogen  atmospheres. Therefore their results are
limited to rather  high  temperature.  Similarly,  \citet{zane00}  studied
fully-ionized hydrogen atmospheres  of accreting  neutron stars. Fully-ionized
hydrogen atmosphere models have been applied to 
 data on isolated neutron stars obtained by {\it XMM-Newton} and {\it
Chandra} \citep{paerels01, pavlov01}. In contrast, radiative transfer in
partially-ionized hydrogen atmosphere with full inclusion of bound-bound and
bound-free opacities remains to be solved in comparison with recent X-ray
data.  

On the  other hand, atmosphere models  of high $Z$  atoms are still immature
due  to their increased  complexity.  \citet{miller92}  was  the first  to
construct model  atmospheres for $Z=2-7$. However, the  opacities include only
bound-free  cross  sections  with  averaging  of the  polarization  modes.  Fe
atmospheres  were investigated  by  \citet{rajagopal97}; the  model is  rather
crude and the energy values and  oscillator strengths have as much as 10\% and
a factor of two uncertainties, respectively. Both of the atmosphere models for
$Z>1$ elements are  based on atomic calculation by  the Hartree-Fock method of
\citet{neuhauser87}.

As the spectroscopy capability of X-ray telescopes improves, model atmospheres
of strongly-magnetized  neutron stars  should comparably improve.  Our efforts
are  focused  on  constructing  atmosphere  models  for  $Z>1$  atoms  for 
comparison  with  existing  hydrogen   atmosphere  models  and  new  X-ray
observations. As part of this effort,  we present  a  novel atomic  calculation
appropriate  for model  atmospheres  of strongly-magnetized  neutron stars.  A
brief review of atomic structure  in strong magnetic fields is given in \S
3, then various  atomic calculations are described in \S4 in comparison with  our formalism in  \S5.
Subsequently, the validity of our model and relevant physical
processes are discussed  in \S 6. Some numerical  improvements are mentioned in
\S7. Finally,  we present  our results  in comparison  with previous atomic models in \S 8. 

% -------------------------------------------------------------------------

\section{Atomic structure in strong magnetic field}
\label{atom_struct}

When a strong magnetic field is  present, the spherical symmetry of an atom is
broken and the atom is stretched along the field. The Landau regime is defined
as  that where  the magnetic  field effects  exceed Coulomb  field effects,
i.e.  $\hbar\omega_B  >  Ze^2/r$  ($\hbar\omega_B=11.57B_{12} \mbox{  keV}=$
electron cyclotron energy, $B=10^{12}B_{12} \mbox{ G}$). This is translated to
$\beta_Z  \ge 1$,  where $\beta_Z=B/Z^2B_0$,  $\beta=B/B_0$ and  the reference
field  $B_0=4.701\times10^9 \mbox{  G}$, as  a good  measure for  the relative
strength of magnetic field to the  nuclear Coulomb field. The dominance of the
magnetic  field requires  us to  investigate the  atomic  structure exploiting
cylindrical  symmetry. In  cylindrical  coordinates $(\rho,  \phi, z)$  bound
electrons in  an atom are well described  by the four quantum  numbers $(n, m,
\nu,  s_z)$ denoting  Landau  number, magnetic  quantum number,  longitudinal
quantum   number   and  component   of   electron   spin   along  the   field
respectively. $\nu$  is the node  of the longitudinal wavefunction and  it defines
the parity along  the field as $(-1)^{\nu}$. $\nu=0$  states are located close
to the nucleus  (\tb orbitals), while $\nu>0$ states are  far from the nucleus
(\lb orbitals).

In  very  strong magnetic  fields  ($\beta_Z  \gg 1$) it  is  a fairly  good
approximation  to fix  $n=0$ (adiabatic  approximation) and  $m\le0, s_z=-1/2$
(full-spin-polarization,  FSP hereafter). Various  atomic models  have adopted both
assumptions (i.e. restrict  the two quantum numbers $n$ and  $s_z$) because
they simplify calculations.  Accordingly, bound electrons are  characterized by a
set  of two  quantum  numbers  $(m,\nu)$ ($m$  indicates  its absolute  value,
hereafter).  However, as $\beta_Z$  approaches 1 the occupation  in $n  > 0$
Landau  levels is  not  negligible (configuration  mixing),  so the  adiabatic
approximation does not provide accurate atomic data \citep{jones_md99_1}. With
further decrease in the magnetic  field, spin-flip transitions take place,
i.e. the FSP structure is no longer  a ground state since an electron with the
opposite spin lowers the total energy \citep{ivanov00}.

The intermediate field  regime is defined as that at  which the electric field
and magnetic field becomes comparable. Bound electrons feel differing electric
fields depending on their distance from the nucleus and the screening of nuclear charge
by  inner  electrons.  A  critical  field strength  $B_c$  which  defines  the
intermediate  regime is  determined  by  equating the  Coulomb  force and  the
Lorentz force. Assuming that the distance of the $(m,0)$ orbital to the nucleus is
$\sim   (2m+1)^{1/2}\hat\rho$  where  $\hat\rho$   is  the   cyclotron  radius
$\hat\rho=2.566\times10^{-10}B_{12}^{-1/2}$ cm and $Z_{eff}$ is the effective charge,   $B_c$    is    given   by
\citep{miller90},
\begin{equation}
B_c=\frac{Z_{eff}^2}{2(2m+1)^3}B_0.
\end{equation}

A choice  of either pure spherical  or cylindrical  basis 
functions is not sufficient for  accurate atomic structure calculations in the intermediate field
regime. Non-separability  of $\rho$ and $z$  in the nuclear  Coulomb term prevents  an analytical  solution  of the  Schr{\"o}dinger  equation even  for
hydrogenic atoms. For  $Z > 1$ atoms, electron-electron  interactions add more
complication to the problem.

% -----------------------------------------------------------------------------

\section{Atomic calculations in the Landau regime}
\label{atom_model}

Various atomic models  have been proposed in the Landau  regime. First of all no  experimental  data is  available  at high  magnetic  fields  $B >  10^{7}$
G. Therefore  atomic data are  highly model-dependent. Hydrogen has 
been investigated by various authors for the last 20 years (see \citet{ruder},
\citet{lai00}  and  references  therein).  Very  accurate  energy  values  and
oscillator  strengths were  obtained  at arbitrary  field  strength with  full
inclusion  of various  effects such  as relativistic  correction \citep{chenz92} and motional
Stark effect \citep{potekhin98-2}.

For $Z  > 1$  atoms, the  number of studies  is significantly  smaller. Atomic
calculations for  multi-electron atoms are  often classified as \it  ab initio
\rm  or  self-consistent  methods.   {\it  Ab  initio}  methods  include  the
restricted   variational   method   \citep{chenh74,  flowers77,   glasser75_1,
glasser75_2,  mueller84} and statistical  methods. The  restricted variational
method parameterizes  the wavefunctions  in an  \it ab initio  \rm way,  so the
accuracy  usually  remains  poor  compared  with  other  methods.  Statistical
approaches   such    as   the   Thomas-Fermi   method    (TF   hereafter)   or
Thomas-Fermi-Dirac    method    (TFD    hereafter)    \citep{skjervold84,lieb,
roegnvaldsson,  thorolfsson} are applicable  to an  atom with  many electrons,
usually for $Z>10$.

In general  self-consistent methods  are more accurate  than {\it  ab initio}
methods,  but they  are computationally intensive since  they  entail iterative
calculations.  The one-dimensional  Hartree-Fock method of \citet{neuhauser87}
(1DHF hereafter) was the first self-consistent method for multi-electron atoms
in the  Landau regime. Since the  adiabatic approximation is  assumed in their
calculations, the Hartree-Fock equation  reduces to a one-dimensional equation
along the  magnetic field.  They provided energy  values of ground  states for
$Z=1-18$  and  26.  \citet{miller91}  obtained energy  values  and  oscillator
strengths  for $Z\le 14$  based on  Neuhauser's atomic  code. As  mentioned by
\citet{miller91},  the inclusion of  $n>0$ Landau  levels to  the Hartree-Fock
equation is extremely difficult.

In recent years several sophisticated atomic models were proposed and studied for
 low $Z$  atoms \citep{jones_md99_1,ivanov98, ivanov99,ivanov00}.  One of them
 is the two-dimensional Hartree-Fock method (2DHF hereafter), which unrestricted
 wavefunctions   both   in   the   transverse   and   longitudinal   direction
 \citep{ivanov00}. It unrestricted the  electron spin as well. Therefore, 2DHF
 is in  principle applicable  at any field  strength, providing  very accurate
 results through  improved numerical accuracy.  \citet{jones_md99_1}, who also
 adopted  a  similar   unrestricted  Hartree-Fock  method  with  even-tempered
 gaussian  basis   functions  (UHF  hereafter),  pointed   out  that the  adiabatic
 approximation provides  accurate atomic  data only at  $\beta_Z >  50$. Since
 typical magnetic  fields of neutron stars can  be at $\beta_Z <  50$ for high
 $Z$  surface elements,  energy  values  obtained by  1DHF  could be  severely
 inaccurate in the  soft X-ray band where high  energy resolution spectroscopy
 is now available. On the other  hand, atomic data from 2DHF and UHF were
 computed  only for  energy levels  of ground  states and  several low-lying
 states of low $Z$  atoms ($Z \le 6$ for UHF, $Z \le 10$  for 2DHF) due to
 computation time limits.  However,  spectroscopy requires  energy
 levels and oscillator strengths for various excited states as well.

The  density-functional method  (DF hereafter)  was applied  to multi-electron
atoms  in  the Landau  regime  \citep{jones_pb85_1, jones_pb85_2,  jones_pb86,
koessl88, relovsky,  johnsen, holas97}. It parameterizes the  relevant terms in
the Hamiltonian (including  the exchange term and electron correlation  term) as a
function of electron  density in the Kohn-Sham equation.  However, results are
highly dependent on the  choice of the exchange-correlation density functional
\citep{jones_pb85_1}. \citet{neuhauser87} mentioned that the density functionals of
\citet{jones_pb85_1,  jones_pb86} and \citet{koessl88} are  not  quite  correct in  high
magnetic fields.  In addition,  DF becomes  inaccurate for an  atom with  a few
electrons. 

Other  atomic calculations with accuracy $\sim10^{-4}$  such as the
Quantum Monte-Carlo (QMC) method  or the Configuration Interaction (CI) method
have been  proposed but  have been implemented  only for  two-electron systems
\citep{jones_md97,  scrinzi,becken99,  becken00}. CI method has  been extended
to high angular momentum states  of  helium \citep{becken01}.  Nevertheless,
it  required   9  months  of computing  relevant matrix elements  to produce
an exhaustive  atomic database. Recently, spectral features from
the white dwarf GD229 have been successfully explained by     helium     in a
magnetic field of  few$\times10^8$     G     \citep{jordan98, jones_md99_2}.

Given the lack of experimental data, the self-consistent method is required to
generate atomic data sufficiently accurate for refined X-ray spectroscopy. An
extensive and accurate atomic data set for various atoms and ion states from
$Z=1$ to 26 in the Landau regime is necessary for the X-ray spectroscopy of
cooling neutron stars.  This contrasts with recent work which has
concentrated on low $Z$ atoms spanning a wide range of magnetic field, achieved
with 2DHF and UHF. 

In the next section, after a brief description of our formalism, we describe our novel approach, which we call the multi-configurational, perturbative, hybrid, Hartree, Hartree-Fock method (MCPH$^3$).  The technique allows us to achieve an exhaustive list of energy values and oscillator strengths at various magnetic fields with very fast computation times on small computers.

% -----------------------------------------------------------------------------

\section{Formalism}
\label{formalism}

\subsection{Schr{\"o}dinger equation preliminaries}

The non-relativistic Hamiltonian in a uniform magnetic field (assuming
infinite nuclear mass) $H$ is divided into a 0th order Hamiltonian $H^{(0)}$
and a perturbation $H^{(1)}$. We set the electron g--factor to 2. 
\begin{eqnarray}
\label{5.1}
H&=&\sum_{i}\left[\frac{1}{2m_e}\left(\vec{p}_i+\frac{e\vec{A}}{c}\right)^2+\frac{e}{m_ec}\vec{B}\cdot\vec{s}_i-\frac{Ze^2}{|\vec{r}_i|}\right]\nonumber\\
& &+\sum_{i}\sum_{j<i}\frac{e^2}{|\vec{r}_i-\vec{r}_j|}\\
\nonumber\\
\label{5.2}
&=&\sum_{i}\Bigg[\frac{1}{2m_e}\left(\vec{p}_i+\frac{e\vec{A}}{c}\right)^2+\frac{e}{m_ec}\vec{B}\cdot\vec{s}_i+V_{nuc}(\vec{r}_i)\nonumber\\
& &+V_{eff}(\vec{r}_i)\Bigg]+\sum_{i}\left[\left(\sum_{j<i}\frac{e^2}{|\vec{r}_i-\vec{r}_j|}\right)-V_{eff}(\vec{r}_i)\right]\\
\nonumber\\
&=&H^{(0)}+H^{(1)}.\nonumber
%&=&\sum_i h^{(0)}_i(\vec{r})+\sum_i h^{(1)}_i(\vec{r})\nonumber
\end{eqnarray}
The vector potential of the uniform magnetic field
$\vec{A}=\frac{1}{2}\vec{B}\times\vec{r}$ is chosen so that the magnetic field
is along the $z$ coordinate $\vec{B}=\nabla\times \vec{A}=B\hat{z}$. Hereafter,
all the lengths are measured in units of $\hat\rho$. $V_{nuc}(\vec{r}_i)$ is
the nuclear Coulomb potential. $V_{eff}(\vec{r}_i)$ is the effective potential
for electron $i$ and represents 
the mean Coulomb potential from the other electrons via the Hartree potential.
\begin{equation}
\label{5.3}
V_{eff}(\vec{r}_i)=\frac{e^2}{\hat\rho}\sum_{j\ne{i}}\int{d^3\vec{r}_j}\;\frac{|\phi_{m_j\nu_j}(\vec{r}_j)|^2}{|\vec{r}_i-\vec{r}_j|}.
\end{equation}
$H^{(0)}$ consists of single-orbital Hamiltonians : 
\begin{equation}
H^{(0)}=\sum_i h^{(0)}_i.
\end{equation}
For an orbital $(m, \nu)$ (hereafter we omit the index $i$), the
Schr{\"o}dinger equation in the Hartree approximation becomes, 
\begin{equation}
\label{5.4}
h^{(0)}\cdot{\phi_{m\nu}(\vec{r})}=\epsilon^{(0)}_{m\nu}\cdot{\phi_{m\nu}(\vec{r})}. 
\end{equation}

\subsection{Multiconfigurational, Perturbative, Hybrid, Hartree, Hartree-Fock:
The Approach} 
\label{sec5.2}

Given  the  formalism above  we  can now  outline  the  MCPH$^3$ approach  and
describe its advantage.  The MCPH$^3$ method starts with equation (\ref{5.4}),
which is the single-orbital Hartree equation for the unperturbed Hamiltonian
$H^{(0)}$.  We  note that (1) $\beta_Z >  1$ and (2) the  Hartree potential is
orbital  independent. The  former implies  the occupation  of $n  >  0$ Landau
levels is small compared to the  $n=0$ state. Therefore equation (\ref{5.4}) is amenable
to  a  solution  in which  we  break  the  single-orbital Hamiltonian ($h^{(0)}$) into  a
0th order term ($\tilde h^{(0)}$) and a 1st order term ($\tilde h^{(1)}$). 
The $\tilde h^{(0)}$
term  will  represent  a  purely single-configurational  calculation  for  the
single-orbital     wavefunction.     The    orbital-independence     of    the
Hartree-potential will  ensure that we can restrict  the iterative calculation
of  the   Hartree  wavefunction  to  the   $n=0$  single-configurational  (one
dimensional) calculation;
once  we   have  determined  the   longitudinal  part  of   this  wavefunction
($f_{m\nu}$)   iteratively,  it   will  be   immediately  applicable   to  the
construction of the multi-configurational Hartree single-orbital wavefunctions
without further  need for iterative calculations. This  represents an enormous
computational  simplification over  the  schemes employed  for hydrogen  atoms
\citep{roesner84, forster84} in which the longitudinal wavefunction for each Landau level in the multi-configurational solution must
be calculated separately.

With   a  set   of   eigenfunctions  ($\chi_{nm\nu}$)   for   the  0th   order
single-configurational Hamiltonian $\tilde h^{(0)}$ determined we can solve for the
multi-configurational single-orbital wavefunctions ($\phi_{m\nu}$) and energy eigenvalues ($\epsilon^{(0)}_{m\nu}$)
using  perturbation theory  with  $\tilde h^{(1)}$.   We  construct the  total
wavefunction   $\Psi(\vec{r})$   for  $H^{(0)}$   by   forming  a   completely
antisymmetrized spatial wavefunction from the $\phi_{m\nu}$ (as required by
the FSP assumption). This total wavefunction $\Psi(\vec{r})$ can then be used
along with $H^{(1)}$ of equation (\ref{5.2}) to solve for the energy of the total Hamiltonian using simple
1st order perturbation theory. 

This approach is, in effect, a double application of perturbation theory. In
the first application perturbation theory is used on the
single-orbital Hamiltonian to handle the higher Landau levels and solve for
the multi-configurational, single-orbital energies and wavefunctions (with
only 1 iterative calculation required). In the second application 1st order
perturbation theory on the total wavefunction and total Hamiltonian yields the
total energy. This same wavefunction is used to calculate oscillator
strengths. To avoid confusion we call the first application
perturbation theory of type I and the second application perturbation theory
of type II.  

\begin{figure}
\epsscale{0.5}
\caption{Comparison of various atomic models in the phase diagram of 
$Z$ and $\beta_Z$. 1DHF = Hartree-Fock method with the adiabatic
approximation \citep{neuhauser86}. 2DHF = Two-dimensional Hartree-Fock method \citep{ivanov00}. DF = Density functional method. TFD = Thomas-Fermi-Dirac method. \label{regime}}
\plotone{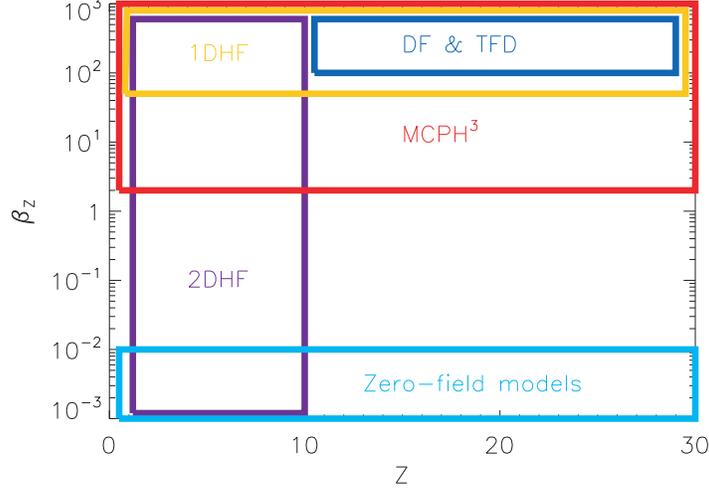}
\end{figure}

This is similar to the Z--expansion method in which the whole electron-electron
interaction  term is  treated as  a perturbation  \citep{hylleras30}. However,
 the Z--expansion method  is accurate only for  highly ionized atoms.  Our 0th order
\wf   is  much   more   accurate   for  multi-electron   atoms   due  to   its
self-consistency. As a result,  we achieved fast convergence (3--5 iterations
for less  than 0.1\%  convergence in total  energies) and  numerical stability
over all the electron configurations  we have computed for $Z=1-26$ atoms. The
method  presented in  this article  is valid  to $\beta_Z  > 1$,  covering the
typical magnetic fields  of cooling neutron stars for  the surface elements up
to Fe (figure \ref{regime}).

\subsection{The MCPH$^3$ solution for the multiconfigurational, single-orbital Hartree wavefunctions and energies (Type I perturbation theory)}

The single-orbital Hamiltonian $h^{(0)}$ in (\ref{5.4}) is divided
into the 0th order and its perturbation as follows. 
\begin{equation}
\label{5.6}
h^{(0)}=\tilde h^{(0)}+\tilde h^{(1)},
\end{equation}
where
\begin{eqnarray}
\label{5.7}
\tilde h^{(0)}&=&h_B(\rho, \phi)+h_L(z),\\
\label{5.8}
\tilde h^{(1)}&=&\{V_{nuc}(\vec{r})-V_{nuc}(z)\}\nonumber\\
& &+\{V_{eff}(\vec{r})-V_{eff}(z)\}.
\end{eqnarray}
$h_B(\rho, \phi)$ is the transverse Hamiltonian, whose eigenvalue and
eigenfunction are $n\hbar\omega_B$ and $\Phi_{nm}(\rho,\phi)$ (see below). Due
to the separation of variables in $\tilde h^{(0)}$, eigenfunctions and
eigenvalues of the Schr{\"o}dinger equation for $\tilde h^{(0)}$ :
\begin{equation}
\label{5.9}
\tilde h^{(0)}\cdot\chi_{nm\nu}(\vec{r})=\epsilon_{nm\nu}\cdot\chi_{nm\nu}(\vec{r})
\end{equation}
are given by 
\begin{eqnarray}
\label{5.10}
\chi_{nm\nu}(\vec{r})&=&\Phi_{nm}(\rho,\phi)\cdot{f_{m\nu}(z)},\\
\label{5.11}
\epsilon_{nm\nu}&=&n\hbar\omega_B+\epsilon_{0m\nu}.
\end{eqnarray}
\{$\chi_{nm\nu}$\} is a complete set of orthogonal
eigenfunctions of $\tilde h^{(0)}$. $\epsilon_{0m\nu}$ is an energy eigenvalue obtained by the
single-configurational calculation for $n=0$ (see below). The
transverse \wf $\Phi_{nm}(\rho, \phi)$ is a Landau function defined as \citep{landau65}, 
\begin{eqnarray}
\label{5.12}
\Phi_{nm}(\rho,\phi)=\frac{\sqrt{n!}}{\sqrt{2\pi(n+m)!}}\left({\frac{\rho}{\sqrt{2}}}\right)^{m}e^{-\rho^2/4}\nonumber\\
\times L^{m}_n\left(\frac{\rho^2}{2}\right)e^{-im\phi},
\end{eqnarray}
satisfying the orthonormal condition
\begin{equation}
\int{\rho d\rho}\int{d\phi}\;\Phi^*_{nm}(\rho,\phi)\cdot\Phi_{n'm'}(\rho,\phi)=\delta_{nn'}\delta_{mm'}.
\end{equation}

\subsubsection{Single-configurational calculation for $n=0$}

An equation for $h_L(z)$ can be found by multiplying the equation (\ref{5.4})
by $\Phi^*_{0m}(\rho,\phi)$ and integrating over $\rho$ and $\phi$. 
\begin{equation}
\label{5.13}
h_L(z)\cdot{f_{m\nu}(z)}=\epsilon_{0m\nu}\cdot{f_{m\nu}(z)},
\end{equation}
where
\begin{equation}
\label{5.14}
h_L(z)=-\frac{\hbar^2}{2m_e\hat\rho^2}\frac{d^2}{d{z}^2}+V_{nuc}(z)+V_{eff}(z).
\end{equation}

The single-configurational nuclear and effective potentials are defined for
$n=0$ as, 
\begin{eqnarray}
\label{5.15}
V_{nuc}(z)&=&\int{\rho{d\rho}}\int{d\phi}\;V_{nuc}(\rho,z)|\Phi_{0m}(\rho,\phi)|^2,
\nonumber\\
\label{5.16}
V_{eff}(z)&=&\int{\rho{d\rho}}\int{d\phi}\;V_{eff}(\rho,\phi,z)|\Phi_{0m}(\rho,\phi)|^2.
\end{eqnarray}

We seek an eigenvalue $\epsilon_{0m\nu}$ with the node $\nu$ under the
boundary condition $f_{m\nu}(z)\rightarrow 0$ at $z=\pm\infty $ with the
solution for $f_{m\nu}(z)$. The only iterative part to the entire solution for
$H$ is complete with this calculation toward the convergence of
$\epsilon_{0m\nu}$ and $f_{m\nu}(z)$; the one-dimensional effective potential
$V_{eff}(z)$ is constructed from wavefunction of other orbitals and the
equation (\ref{5.13}) is solved for a new energy eigenvalue and longitudinal wavefunction. 

\subsubsection{Perturbation to higher Landau levels}

With the eigenvalues and eigenfunctions of $\tilde h^{(0)}$ determined, we can
proceed to obtain the multi-configurational, single-orbital wavefunction
$\phi_{m\nu}(\vec{r})$ using perturbation theory. We expand
$\phi_{m\nu}(\vec{r})$ as 
\begin{equation}
\label{5.17}
\phi_{m\nu}(\vec{r})=\sum_{n}c_{nm\nu}\cdot\chi_{nm\nu}(\vec{r}). 
\end{equation}
The coefficients $c_{nm\nu}$ are computed by the usual non-degenerate
perturbation theory (as is the energy $\epsilon^{(0)}_{m\nu}$).
\begin{equation}
\label{5.18}
c_{nm\nu}=c^{(0)}_{nm\nu}+c^{(1)}_{nm\nu}+c^{(2)}_{nm\nu}+...
\end{equation}

Matrix elements of the 0th order Hamiltonian between different $n$ but
the same $(m,\nu)$ are simply $\langle
nm\nu|\tilde h^{(0)}|n'm\nu\rangle=(n'-n)\hbar\omega_B$ where $\langle \vec{r}|nm\nu\rangle\equiv\chi_{nm\nu}(\vec{r})$. The
matrix elements of $\tilde h^{(0)}$ become 0 for different $m$ states
due to the orthogonality of Landau functions. The perturbation terms between
different $\nu$ states are also 0, because different $\nu$ states are
orthogonal to each other. On the other hand, matrix elements of
$\tilde h^{(1)}$ between different Landau states consist of two parts as follows. 
\begin{eqnarray}
\label{5.19}
\langle nm\nu|\tilde h^{(1)}|n'm\nu\rangle&=&\langle{nm\nu}|V_{nuc}(\vec{r})|n'm\nu\rangle\nonumber\\
& &+\langle{nm\nu}|V_{eff}(\vec{r})|n'm\nu\rangle.
\end{eqnarray}
For the evaluation of the matrix elements, the reader should refer to the
appendix. 

\subsection{The MCPH$^3$ solution for the total wavefunction and energy of the Hamiltonian (Type II perturbation theory)}

Once the Hartree single-orbital wavefunctions ${\phi_{m\nu}(\vec{r})}$ in (\ref{5.17}) are
orthonormalized, the 0th order total wavefunction is given as below since the 
spatial wavefunction must be antisymmetrized based on the FSP assumption. 

\begin{equation}
\label{5.20}
\Psi(\vec{r})=A\prod_{m\nu}{\phi_{m\nu}}(\vec{r}),
\end{equation}
where $A$ is the antisymmetrizing operator. 

Based on the total spatial wavefunction in (\ref{5.20}), we evaluate the
single-orbital energy (which is the ionization threshold) $\epsilon_{m\nu}$ and total energy $E$ to the 1st
order in perturbation theory (called type II perturbation theory in
\S\ref{sec5.2}). The single-orbital energy is given by,  
\begin{equation}
\label{5.21}
\epsilon_{m\nu}=\epsilon^{(0)}_{m\nu}+\epsilon^{(1)}_{m\nu},
\end{equation}
where 
\begin{eqnarray}
\label{5.22}
\epsilon^{(1)}_{m\nu}&=&\langle \Psi |
\left(\sum_{m'\nu'}\frac{e^2}{|\vec{r}-\vec{r}\,'|} -V_{eff}(\vec{r})\right) | \Psi \rangle \nonumber\\
&=&-\frac{e^2}{\hat\rho}\sum_{m'\nu'}\int{d^3\vec{r}}\int{d^3\vec{r}\,'}\phi^*_{m\nu}(\vec{r}\,')\phi^*_{m'\nu'}(\vec{r})\frac{1}{|\vec{r}-\vec{r}\,'|}\nonumber\\
& &\times\phi_{m\nu}(\vec{r})\phi_{m'\nu'}(\vec{r}\,').
\end{eqnarray}
Only the exchange term remains in 1st order because the direct Coulomb term
and the effective potential cancel out when the Hartree single-orbital
wavefunctions have converged in the iteration associated with (\ref{5.13}).

Total energy $E$ to 1st order is given by, 
\begin{eqnarray}
\label{5.23}
E&=&\sum_{m\nu}{\epsilon^{(0)}_{m\nu}}+E^{(1)},\\
\nonumber\\
E^{(1)}&=&\langle\Psi|H^{(1)}|\Psi\rangle\nonumber\\
&=&-\frac{e^2}{2\hat\rho}\;\sum_{m\nu}\sum_{m'\nu'}\int{d^3\vec{r}}\int{d^3\vec{r}\,'}\nonumber\\
& &\Bigg(|\phi_{m\nu}(\vec{r})|^2|\phi_{m'\nu'}(\vec{r}\,')|^2\frac{1}{|\vec{r}-\vec{r}\,'|}\nonumber\\
& &+\phi^*_{m\nu}(\vec{r}\,')\phi^*_{m'\nu'}(\vec{r})\frac{1}{|\vec{r}-\vec{r}\,'|}\phi_{m\nu}(\vec{r})\phi_{m'\nu'}(\vec{r}\,')\Bigg).
\end{eqnarray}
In contrast, there remains part of the direct Coulomb term in $E^{(1)}$
following the way we have divided the Hamiltonian $H$ into $H^{(0)}$ and
$H^{(1)}$ in (\ref{5.2}). However the applicability of type II perturbation is
determined by the ratio $|\epsilon^{(1)}_{m\nu}/\epsilon^{(0)}_{m\nu}|$ as is
discussed in \S\ref{exchange}. 

\subsection{Oscillator strengths}

Oscillator strength in the length form for the transition $\Delta{M}=q$ is
given as ($M$ is the total magnetic quantum number $M=\sum_j m_j$),
\begin{equation}
f^{(q)}_{fi}=\frac{2m_e\hat\rho^2E_{fi}}{\hbar^2}|\langle{f}|\sum_{j}{r_{j}^{(q)}}|i\rangle|^2,\\
\end{equation}
where
\begin{equation}
r_{j}^{(q)}=\sqrt{\frac{4\pi}{3}}r_jY_{1q}(\theta_j,\phi_j).
\end{equation}
$Y_{lq}$ is the spherical harmonic. $E_{fi}$ is the transition energy. Selection rule for the dipole-allowed
 transitions is :
\begin{eqnarray*}
& &\Delta m=\pm1,\;\; \Delta\nu=even\\
& &\Delta m=0,\;\;\;\;\; \Delta \nu=odd
\end{eqnarray*}

\subsubsection{Dipole matrices}

The dipole matrix elements $d^{(q)}$ in units of $\hat\rho$ are defined as,
\begin{equation}
d^{(q)}_{m\nu,m'\nu'}{\equiv}\langle{m'\nu'|r^{(q)}|m\nu}\rangle.
\end{equation}
$|m\nu\rangle$ and $|m'\nu'\rangle$ denote an orbital in initial and final
state respectively. Using the orthonormality of Landau functions, dipole
matrix elements are evaluated as, 
\begin{eqnarray}
d^{(0)}_{m\nu,m'\nu'}&=&\sum_{n,
n'}c_{nm\nu}c_{n'm'\nu'}\delta_{m'm}\delta_{n'n}\nonumber\\
& &\times\int{dz}\;{f^*_{m'\nu'}(z)\,{z}\,f_{m\nu}(z)},\\
\nonumber\\
d^{(+1)}_{m\nu,m'\nu'}&=&-\sum_{n,
n'}c_{nm\nu}c_{n'm'\nu'}\delta_{m'm+1}[\delta_{n'n+1}\sqrt{n+1}\nonumber\\
& &+\delta_{n'n}\sqrt{n+m}]\int{dz}\;{f^*_{m'\nu'}(z)f_{m\nu}(z)},\\
\nonumber\\
d^{(-1)}_{m\nu,m'\nu'}&=&-\sum_{n,
n'}c_{nm\nu}c_{n'm'\nu'}\delta_{m'm-1}[\delta_{n'n-1}\sqrt{n}\nonumber\\
&+&\delta_{n'n}\sqrt{n+m+1}]\int{dz}\;{f^*_{m'\nu'}(z)f_{m\nu}(z)}.
\end{eqnarray}

The longitudinal wavefunction for the initial and the final states are not
quite orthogonal to each other due to differing electron configurations.
Nevertheless we evaluated oscillator strengths for the single electron
transitions using only the dipole matrix element of the transitioning
electron.  The error caused by this truncation is insignificant for one
electron transitions. 

% --------------------------------------------------------------------------

\section{Validity and other physical effects}
\label{validity}

\subsection{Valid region in $B$--$Z$ phase space}

\begin{figure}
\epsscale{0.5}
\caption{Critical magnetic field $B_c$ for a \tb orbital $(m, 0)$ and the
spin-flip transition field strength $B_{sf}$. We assumed $Z_{eff}\simeq Z-m$
for illustrative purpose. We did a polynomial fit to the
spin-flip field strengths obtained by \citet{ivanov00}. $B_{sf}=1.7\times10^{9}$ G for $Z=2$, $B_{sf}=9.6\times10^{10}$ G for
$Z=10$ and $B_{sf}=6.8\times10^{11}$ G for $Z=26$. Triangle points represent
$B_{sf}$ for neutral atoms, while crossed points are for singly-ionized
atoms. \label{spinflip}}
\plotone{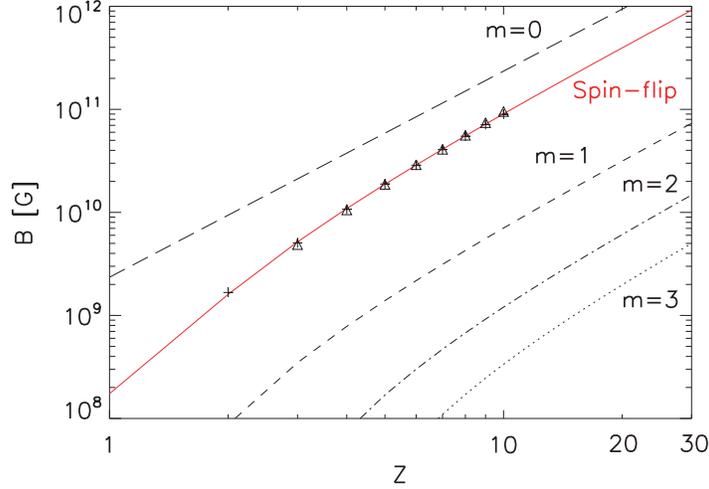}
\end{figure}

An expansion  of the single-particle Hamiltonian in terms of cylindrical
wavefunctions provides  accurate  results  when $B>B_c$. As mentioned in
\S\ref{atom_struct} this is a statement that magnetic field effects dominate over Coulomb
field effects. On  the other  hand, the perturbative
treatment of  higher Landau levels is valid when $|\tilde h^{(1)}| < |\tilde h^{(0)}|$. Since the
nuclear  term in  equation (\ref{5.19}) dominates  the effective  potential  term, the 
condition $B>B_c$ is also sufficient to ensure the validity of a perturbative
treatment of the higher Landau levels. Therefore, $B_c$  sets a lower limit on
the magnetic field for which our method  is valid for \tb  electrons (i.e. in
state $(m,0)$). In figure \ref{spinflip},  we plot $B_c$ for  $m=0-3$ assuming $Z_{eff}\simeq
Z-m$  for illustrative purposes.  Loosely-bound electrons  are well  within the
Landau regime due to their  larger average distance to the nucleus. Therefore
even  at intermediate  field strengths  a  cylindrical expansion  combined
with perturbative treatment of the higher Landau levels still provides accurate results
for most electrons. The sharp decrease in $B_c$ with increasing $m$, seen
in figure \ref{spinflip}, means that only those \tb inner electrons with
$m=0-3$ will have significant configuration mixing. At higher quantum state
$m$, the Landau regime, with its negligible configuration mixing, is quickly
recovered. 

\subsection{Spin-flip transitions}

As the magnetic field gets smaller, spin-flip transitions can  take  place, obviating the
validity of the FSP approximation. The  critical
magnetic field strengths where  spin-flip transitions take place ($B_{sf}$) were recently
studied for atoms up to $Z=10$  by \citet{ivanov00}. In a fairly wide range of
magnetic field  they  computed  the ground  state  energies  of atoms in which the
FSP approximation was removed -- the electron spins were unrestricted. Using
this study we  found  that $B_{sf}$ could be fit by a polynomial function of $Z$. At  $B<B_{sf}$ the assumption of antisymmetrized  total spatial
wave function breaks  down. This sets another restriction  on the validity of
our  method  in  the  phase  space   of  $(B, Z)$ (figure \ref{spinflip}).

\subsection{Exchange term}
\label{exchange}

Although the significance of the exchange term in the binding energy increases
in  the  FSP approximation  \citep{schmelcher99},  the  exchange  term can  be
treated perturbatively  as long  as it  is smaller than  the 0th  order energy
eigenvalues,   i.e.   $|\epsilon^{(1)}_{m\nu}|<|\epsilon^{(0)}_{m\nu}|$.    We
observed  this inequality  was satisfied  for any  single orbital  in  all the
configurations computed. In  the perturbation for the exchange  term 
we  did  not  perform  calculations higher than  0th  and 1st order for  the
wavefunctions and  energy values because  of the increasing complexity  of the
calculations. Also  the naive  inclusion of a 1st  order perturbation  into the
wavefunctions breaks  the wavefunctions orthogonality,  requiring a further
orthogonalization process for the single-orbital wavefunctions. The exact
value of the wavefunction is  not our
primary concern, but  rather the accuracy of the real physical observables such  as energy values and
oscillator strengths. For instance, errors  on the energy levels  caused by
truncating the wavefunctions are actually small (\S\ref{energy}). We use an explicit
procedure to quantitatively correct the oscillator strengths in order to
account for the  deviation  introduced by using  Hartree rather than Hartree-Fock wavefunctions  (\S\ref{os}). 

\subsection{Electron correlation}

In reality the individual wavefunctions are not symmetric but distorted by the Coulomb coupling
with  other   electrons.  The Hartree  and   Hartree-Fock  methods do not
include electron correlation  in  the  Schr{\"o}dinger  equation  for  each
orbital. However,  as the magnetic field increases, the  electron correlation
becomes  less  relevant \citep{schmelcher99}.  The  estimated  error in our
energies and oscillator strengths due to electron correlation is significantly smaller than 1\% in the Landau regime.

\subsection{Relativistic effects}

The significance  of relativistic effects is proportional  to $\hbar\omega_B /
m_ec^2$  or  $E_B /  m_ec^2$  where  $E_B$ is  a  binding  energy. The  former
condition translates to $B/B_{rel}$ ($B_{rel}=4.414\times10^{13} \mbox { G}$),
representing the  degree of relativistic  effects in the transverse  motion of
electrons. However, the  shape of the Landau wavefuntion  in the relativistic
theory  is the same  as in  the non-relativistic  theory \citep{lai00}.  On the
other hand,  the electron becomes  relativistic along the magnetic  field when
$E_B\sim  m_ec^2$. However, the  relativistic effects  on the  Coulomb binding
energies  remain  negligible  even  at  magnetic  field  strengths  of  up  to
$4.7\times10^{13}$ G as determined by numerically solving Dirac's equation for
hydrogen \citep{lindgren79, chenz92}.  Also, the second  condition is relevant  only to
the inner orbitals which are  not important for X-ray spectroscopy since their
binding energies are high ($>$ 10 keV).

\subsection{Finite nuclear mass effects}

In the presence of a magnetic field, the collective motion of an atom and its 
internal degrees of freedom are coupled, and this coupling modifies the
electronics structure of the atom \citep{pavlov93,  potekhin94}. Consequences of
this coupling include
distortion of the wavefunctions, violation of some dipole selection rules and
line broadening. Most work  on these finite nuclear
mass effects has been  performed on
hydrogen atoms \citep{lai92, pavlov93, potekhin94, kopidakis96, potekhin98-2} and hydrogenic ions \citep{bezchastnov98}. 

The immediate impact on our work is that the energy levels and oscillator
strengths are affected by these collective effects. In this section we briefly
review the collective motion and the theoretical framework for addressing
it. In this paper we do not relax the assumption of an infinite nuclear mass
in the Hamiltonian. This facilitates direct numeric comparison of our MCPH$^3$
approach with other approaches for solving the strong magnetic field Hamiltonian,
since they also assumed infinite nuclear mass. We do explain, in light of
previous work on hydrogenic systems, how we will incorporate the collective
motion into the MCPH$^3$ technique in a subsequent paper. In the case of the
bound-bound transitions of interest to us we describe the straightforward
incorporation of finite nuclear mass into the oscillator strength
calculation. This is appropriate since previous hydrogen work also included
the effects of the finite nuclear mass on the oscillator strengths. 

The finite nuclear mass treatment is quite complicated in hydrogen and has not
really be dealt with previously for higher $Z$ atoms. We outline in this section
how substantial simplifications of this problem arise when considering both
the specifics of higher $Z$ systems in general and the requirements of X-ray
spectroscopic modeling in particular. 

\subsubsection{Review of previous approaches to finite nuclear mass effects}

When the nuclear mass is not assumed to be infinite the Hamiltonian, including
a term for nuclear motion, can be separated into center-of-mass (CM) and
relative coordinates by a canonical  transformation, yielding 
\begin{equation}
H\rightarrow H'=H_{cm}+H_{int}+H_{c},
\end{equation}
where $H_{cm}$  and $H_{int}$ are  the Hamiltonian for  the CM motion  and the
internal part. There is a scaling law for $H_{int}$ from the infinite
nuclear mass case (corresponding to the Hamiltonian in equation (\ref{5.1})) to the
finite mass case \citep{pavlov-verevkin80-1, becken99}.  Even when there is no
translation motion of the atom, there is an additional term $m\hbar\Omega_B$ ($m$ is the $z$-component of angular
momentum and  $\Omega_B$ is  the cyclotron frequency  of the  nucleus) arising
from the cyclotron motion of the nucleus (hereafter we call this term as the
nuclear cyclotron term). Due to the nuclear cyclotron term, the large $m$
states are autoionized \citep{kopidakis96}. Since this term commutes with the Hamiltonian, it is easy to
implement as an additive term in $H_{int}$. Less straightforward to handle is
the coupling term $H_c$. This term arises because the motion of the CM through the magnetic field induces an electric field -- the so-called motional Stark
effect. The motional  Stark field acts as a dipole electric
field  which increases the separation between  electron and nucleus and
therefore  decreases the binding energy compared to the infinite nuclear mass
case. $H_c$ can be expressed as :
\begin{equation}
H_c=\frac{e}{Mc}(\vec{K}\times \vec{B})\cdot\vec{r},
\end{equation}
where $M$ is  the nuclear mass and $\vec{r}$ is the relative coordinate between
the nucleus and the electron. $\vec{K} = \vec{\pi} - e(\vec{B}\times\vec{r})/c$
is the pseudomomentum, where $\vec{\pi} = \vec{p}+e\vec{A}/c$ represents the kinetic momentum. The pseudomomentum 
commutes  with the  Hamiltonian and  therefore is  a constant  of  motion.
The pseudomomentum
defines  a separation  between  the guiding  centers  of the  nucleus and  the
electron;  $\vec{r}_c = \case{c}{eB^2}(\vec{K}  \times\vec{B})$.   The
components of  the pseudomomentum commute for  neutral atoms,  while they do
not commute for a charged system \citep{baye90}. 

The motional Stark field breaks  the  cylindrical symmetry  which is present by
the central Coulomb field in the  infinite
nuclear  mass  case. As a result the system observables generally  depend  on
$K_{\perp}$, the transverse component of the pseudomomentum.  For
instance, we express the binding energy of an electron  in an orbital $(m,\nu)$
as $\epsilon_{m\nu}(K_{\perp})$. States can be classified according
to the relative size of $K_{\perp}$ compared to a critical pseudomomentum $K_c
\sim   (2M\epsilon_{m\nu}(0))^{1/2}$  and
$\epsilon_{m\nu}(0)$ is the binding energy of the electron in a non-moving
atom. When $K_{\perp}<K_c$ the central Coulomb field exceeds the Stark field;
this is called a centered state. When $K_{\perp} > K_c$ the Stark field
dominates and it separates the electron from its equilibrium (Coulomb)
position around the nucleus; this is a decentered state. 

Treatment of collective effects in the Hamiltonian depends on whether a state
is centered or decentered. For decentered states a coordinate system $\vec{r}\,' = \vec{r}-\vec{r}_c$ is useful because the
coupling term $H_c$ becomes zero  in the transformed Hamiltonian $H'$. Instead 
the nuclear  Coulomb term is  shifted by $\vec{r}_c$, requiring  evaluation of
matrix elements at each $K_{\perp}$. For nearly centered states the motional
Stark effect can be treated by considering the coupling term $H_c$ as a
perturbation to the infinite nuclear mass solution \citep{pavlov93}. In the perturbation method binding energy is given by,
\begin{equation}
\epsilon_{m\nu}(K_{\perp})=\epsilon_{m\nu}(0)+\frac{K_{\perp}^2}{2M^{\perp}_{m\nu}}.
\end{equation}
$M^{\perp}_{m\nu}$ is  the transverse mass, which can  be obtained by the
2nd  order  perturbation, since  the bound  states are non-degenerate.  The
transverse mass is larger than the atomic mass $M$ (mass anisotropy). 

Non-perturbative calculation was first studied by \citet{vincke92}. \citet{potekhin94} studied binding energies, oscillator strengths and size of
moving hydrogen atom by the multi-configurational method.  \citet{potekhin98-2}
obtained fitting formula based on the results of \citet{potekhin94}.  \citet{bezchastnov98} investigated hydrogen-like  helium by
multi-configurational  method with  two  particle basis  sets.  For a  charged
system the collective motion is   quantized   by  the   cyclotron   motion
due   to  the   non-zero   net charge. Therefore atomic structure is
characterized by the Landau numbers of the
collective cyclotron motion instead of $K_{\perp}$. \citet{baye90} and
\citet{schmelcher95} have investigated a unified picture of general neutral and charged systems. 

\subsubsection{Effect of finite nuclear mass on X-ray spectral features in MCPH$^3$}

We have only considered the infinite nuclear mass case in this paper to
validate MCPH$^3$ against other methods, which did not consider collective
effects in non-hydrogenic systems. However, we included a correction due to
the nuclear cyclotron term for oscillator strengths of $\Delta m \ne0$
transitions for hydrogen (mentioned elsewhere in \S\ref{os}). We outline here
the program we will implement in a subsequent paper to accommodate these
collective effects within the MCPH$^3$ formalism. 

First of all, the degree of motional Stark effect decreases monotonically
with the nuclear mass $M$. In addition, the binding energy of the inner electrons increases
as the nuclear charge $Z$ becomes larger. Therefore, the tightly-bound
states are centered or nearly centered states. This allows us to adopt the
perturbation method for centered states as employed by \citet{pavlov93} and 
\citet{lai95}. This approach is fully compatible with MCPH$^3$ but will set a
constraint on the density range of applicability for a given temperature. In particular this approach
is effective at higher densities where highly excited (decentered) states are
destroyed by quasi-static electric fields  from adjacent atoms
and by collisions with fast free electrons (pressure ionization), or are 
autoionized due to the nuclear cyclotron term. As shown by
\citet{potekhin99} the relative population of centered and decentered states
in neutral hydrogen depends on $B,\rho$ and $T$ in a complicated manner. The
range of validity of the perturbation theory will be the region of parameter
space where centered states dominate. The range of parameter space in which a
perturbative approach is valid will be much larger in the case of higher $Z$
atoms compared to hydrogen because of the heavier nuclear mass involved. We also note that, at sufficiently high $B$, differences in the
nuclear cyclotron term for different $m$ states can be larger than the
distances between single-orbital energy levels. This results in level mixing
and sets constraint on the applicablity of MCPH$^3$. Taking into account these points the parameter space ($B, Z, \rho, T$) of validity will be the
subject of our next paper to construct realistic atmosphere models.

A number of effects have been considered in the literature that are much less
important for X-ray spectroscopy in the {\it XMM-Newton} and {\it Chandra}
energy bands. For instance, as the motional Stark field decreases the binding energy spectral
features are subject to redward line broadening (magnetic broadening,
\citet{pavlov93}). In addition, the Stark field breaks the cylindrical
symmetry opening up new transitions that are forbidden according to the dipole
selection rules for non-moving atom. \citet{pavlov93} classified
spectral features associated with the motional Stark effect as arising from
high or low energy components. High energy components in the soft X-ray band
consist of transitions from states of large binding energy; they are centered
states. On the other hand, the low energy components of \citet{pavlov93}
result from decentered state transitions leading to radiation in the
optical--UV band. Consequently we will be able to apply the perturbation
methods to the magnetic broadening for spectral features in the soft X-ray band.  

%------------------------------------------------------------------------------

\section{Numerics}
\label{numerics}

\subsection{Spatial grid points and computation time}

\begin{figure}
\epsscale{0.5}
\caption{Algorithm of the Multi-configurational Perturbative Hybrid Hartree
Hartree-Fock method. \label{algorithm}}
\plotone{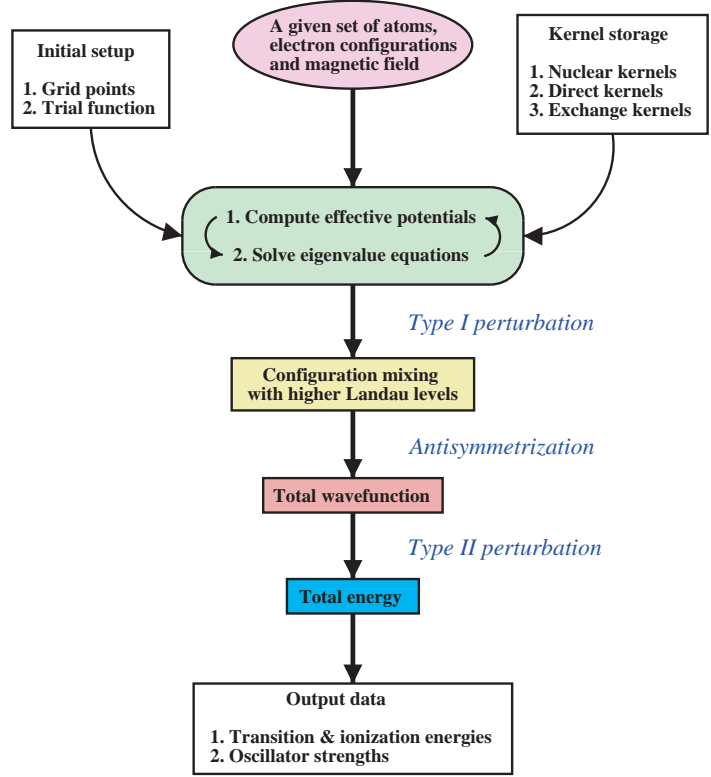}
\end{figure}

The main part of the  iterative calculation is solving the eigenvalue equation
(\ref{5.13}) for   each  orbital  and   the  numerical   evaluation  of   the  relevant
integrals. We  perform a  finite element  method, using a  set of  grid points
\{${\rho_i}$\}  ($N_\rho$   elements,  grid  interval   $\Delta  \rho_i$)  and
\{${z_j}$\} ($N_z$  elements, grid interval  $\Delta z_j$) for  the transverse
and longitudinal direction respectively. Since a $(m, \nu)$ state is localized
around $(2m+1)^{1/2}\hat\rho$ with the  width $\sim\hat\rho$ in the transverse
direction  \citep{meszaros92}, the  transverse size  of $10\hat\rho$  is large
enough  for $Z  < 30$.  We precalculated  all the  relevant  kernels (nuclear,
direct and exchange) as a function  of $z$ by integrating over the grid points
\{$\rho_i$\},  since  the kernels  are  independent of  $B$  when  we use  the
cyclotron radius for  the unit of length. See the  appendix for the evaluation
of  kernels. We  used a  set of  uniform grid  points for  \{${\rho_i}$\} with
$N_\rho=10^3$  and $\Delta \rho_i=10^{-2}$.  We evaluated  the integrals  by a
five-point Newton-Cotes  integration formula. Numerical  error associated with
the kernel integrals is negligibly small. Prior to the calculation for a given
electron configuration,  the kernels are  read from the database  comprised of
about $3\times10^5$  lines of floating-point  numbers. The kernel  database is
extended over $m=0-30$ for  single-configurational calculation and $n=1-7$ for
the    magnetic   quantum    numbers    $m=0-5$   for    multi-configurational
calculation. This is sufficiently large for $Z=1-26$ in the Landau regime. The
algorithm of our atomic calculation is presented in figure \ref{algorithm}.

When  solving an  eigenvalue equation  for each  orbital by  Numerov algorithm
\citep{koonin86}, we confined  the longitudinal coordinate to $0\le  z \le L$,
with  a  sufficiently   large  size  of  the  integration   box  $L$,  because
$f_{m\nu}(z)$ is  symmetric regardless  of the electron  correlation. Boundary
conditions  are :  $f_{m\nu}(z=0)=0$  (Dirichlet condition)  for odd  $\nu$'s,
$f'_{m\nu}(z=0)=0$ for even  $\nu$'s (Neumann condition) and $f_{m\nu}(z=L)=0$
for all $\nu$'s.  In contrast to the transverse  direction, the characteristic
length   of    longitudinal   wavefunctions   varies    greatly   with   $\nu$
\citep{ruderman71}   ($a_0$   is    the   Bohr   radius,   $a_0/{\hat\rho}   =
\sqrt{2}\beta^{1/2}$).
\begin{eqnarray*}
L&\sim&{(a_0/Z\hat\rho)\ln{(a_0/Z\hat\rho)}}\hspace{1cm}(\nu=0)\\
&\sim&{a_0}\nu^2/\hat\rho \hspace{2.8cm}(\nu>0)
\end{eqnarray*}

In compensation  for saving computation  time from the kernel  calculation, we
confront the difficulty of the $B$-dependent size of the integration box as we
keep the  fine grid intervals.  Inner  \tb orbitals require  finer grid points
(especially for high $Z$ atoms), while wavefunctions of outer orbitals require
large extension in integration box, with $L$ as large as $\sim$ 1000 depending
on $B$,  $Z$ and $\nu$.  As the  results are sensitive to  the grid intervals,
uniform grid  points are not  appropriate for an  atom mixed with \tb  and \lb
electrons unless  we have  a large number  of grid points  $N_{grid}$ (meaning
large  computation time).  Therefore,  we adopt  the  non-uniform grid  points
: $z_j=g(\tilde z_j)$ of \citet{ivanov94}. \{$\tilde  z_j$\} is a set of
uniform grid  points on  which  the equation (\ref{5.13}) is  solved, while  \{$z_j$\}  is a  set of  actual
non-uniform grid points.  A function $g(\tilde z)$ must be  chosen so that the
grid intervals $\Delta z_j$ become denser near the origin and sparser at large
$z$.  Upon  the  mapping of  coordinates,  the equation (\ref{5.13})  must  be transformed  into  an
appropriate form  in terms of a new  set of grid points  \{$\tilde z_i$\}. For
some of the \lb  orbitals, we do not know {\it ab  initio} where a \wf changes
steeply,  but  they  do  not   require  finer  grid  points  compared  to  \tb
states. Therefore, we  used uniform grid points with  the interval $\Delta z_j
\sim  0.5-0.8$.  It  reduces  the  number  of  grid  points  significantly  to
$N_z=70-200$ for  \tb orbitals and  $N_z=70-1000$ for \lb orbitals.  Note that
most  electrons are  usually in  \tb  orbitals while  a few  \lb orbitals  are
occupied by excited  states and the ground states of  high $Z$ atoms.  Keeping
the  same  set  of  grid  points for  different  electron  configurations  for
different $B$ and $Z$ is still a difficult task. We adjust the grid points, if
necessary, after the 1st iteration.

Computation   time  is   proportional  to   ${N_{grid}}^2\times  {N_e}^2\times
N_{iter}$  for  the  self-consistent  methods  due  to  the  coupling  between
electrons  ($N_e$ is  the number  of electrons,  $N_{iter}$ is  the  number of
iterations). Use of a one-dimensional Hartree equation is far superior to 2DHF
($N_{grid}=N_\rho\times  N_z=65\times 65=4225$)  \citep{ivanov00} in  terms of
the   computation   time.  \citet{miller91}   adopted   uniform  grid   points
($N_z=500-1000$   and  $\Delta   z=0.2-0.8$)  for   most  of   their  electron
configurations  computed, suffering  the  numerical inaccuracy  caused by  the
coarse grid points, as pointed out by \citet{thurner}.

\subsection{Convergence and numerical errors}

\begin{figure}
\epsscale{0.5}
\caption{Convergence test for the single-orbital Hartree energy of the
innermost orbital $(m,\nu)=(0,0)$ for $Z=2, 6, 14, 26$ neutral
atoms. $B_{12}=1$, $\Delta \tilde{z}=4\times10^{-2}$. All the
electrons are in tightly-bound orbitals. There was minor change in the
convergence speed for different magnetic fields and electron configurations. 
\label{convergence}}
\plotone{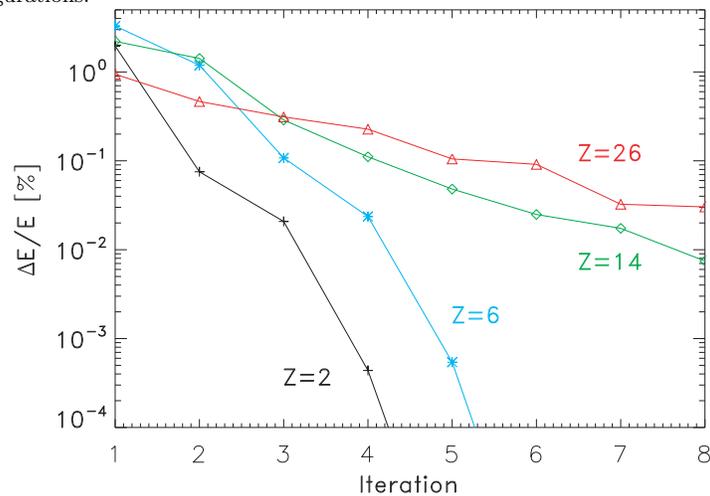}
\end{figure}

Our initial guess  for the longitudinal wavefunction of  the $(m,\nu)$ orbital
is adopted from restricted variational studies \citep{flowers77}.
\begin{equation}
f_{m\nu}(z)\propto{z^{\nu}{e^{-a_{m\nu}|z|}}}\nonumber
\end{equation}
The  final results do  not change  with our  initial choice  of $\{a_{m\nu}\}$
varying from 0.1 to 10. Energy values  converge to less than 1\% after the 2nd
iteration and  less than 0.1\%  after the 4th iteration,  although convergence
speed  becomes slow  as $N_e$  increases (figure \ref{convergence}). Therefore,
estimations of  convergence speed  from hydrogen results  \citep{miller91} are
not correct. Convergence  speed for a 2DHF calculation will  be slower than in
the one-dimensional Hartree method.

Upon  the evaluation  of the  direct  and exchange  terms, we  found that  the
coupling  terms   between  different  $n   >0$  states  are   negligible  ($<$
0.1\%).  Therefore, we  did not  include  the contribution  from $n>0$  Landau
states   of  other   orbitals  (setting   $\tilde{n}  =   \tilde{n}'   =0$  in
(\ref{eq_energy1})  and  (\ref{eq_energy2})  in  the appendix),  reducing  the
computation time  when evaluating  the 1st order  total energy.  The numerical
error is overall less than 0.1\%.
%
% --------------------------------------------------------------------------
%
\section{Results}
\label{results}
\subsection{Energy values and ground state configurations}
\label{energy}

We have computed  total energies for ground states  and several excited states
at  representative  magnetic fields.  The  computed  energies  match with  the
previously most accurate  values (e.g. 2DHF) with less than 1\% deviation. In general,
model-dependent total energy is closer to the exact value when it is lower than
other models. However, the computed energy  values can be lower than the exact
values  due  to  numerical  error  or  inconsistency  in  atomic  models.  At
$\beta_Z$ close to 1,  our energy values are  actually lower  than the  previous results
known to be accurate to 4 digits.  This is because higher Landau states of
the innermost electrons are no longer perturbations on the $n=0$ state. Therefore,
 perturbative  method may  provide total  energies lower than  the previous
accurate data. Nevertheless, we  listed the multi-configurational results as
long as the  discrepancy is  less than  1\%.

\subsubsection{Hydrogen}

We  present the  energies  for ground  states  and several  excited states  in
comparison with the multi-configurational Hartree-Fock method (MCHF hereafter)
\citep{ruder} (table \ref{h_ground_energy}, \ref{h_excite_energy}). They fully
included higher  Landau levels up to  $n=12$ in the  Hartree-Fock equation. We
produced fairly good energy values with $<$ 1\% deviation from MCHF.

\subsubsection{Helium}

We  compare   our  results  with   1DHF  \citep{thurner,  miller91}   and  UHF
\citep{jones_md99_1} for  the ground state  and several excited  states (table
\ref{he_ground_energy},    \ref{he_excite_energy}).     The    ground    state
configuration  for  helium  is  $(m,  \nu)  =  (0,0),  (1,0)$  in  the  Landau
regime.  Our  single-configurational total  energies  (excluding $n>0$  Landau
levels)  match  well with  1DHF  results  \citep{thurner}  by $<$  0.5\%.  Our
multi-configurational total  energies are lower  than 1DHF by $\sim  1-5$\% at
$\beta_Z < 10$ and higher than UHF by $<$ 1\% at $\beta_Z > 1$.

\subsubsection{High Z atoms}

For  $Z>2$   atoms,  published  atomic  data   is  scarce  \citep{neuhauser87,
ivanov00}. First  of all, we searched  for the ground  state configurations of
neutral atoms  $Z=2-26$ at some  representative field strengths [table  5,6]. In
the limit of $B\rightarrow\infty$, the ground state configuration is such that
all  the  bound  electrons are  in  \tb  orbitals  $(m,  \nu) =  (0,0),  (1,0)
...  (N_e-1,0)$.  With decreasing  $B$,  the  spatial  crossover takes  place,
i.e.  having electrons  in \lb  orbitals rather  than in  outer  \tb orbitals,
lowers  the  total  energy.  A  $\nu=2$  orbital may  form  the  ground  state
configuration  because a $\nu=2$  orbital can  be closer  to the  nucleus than
$\nu=1$ orbitals due to  the electron screening \citep{jones_pb85_2}. However,
we did not  find any ground state configurations with  $\nu=2$ orbitals at the
magnetic       fields      in       tables       \ref{ground_energy1}      and
\ref{ground_energy2}.  Several  spatial   crossovers  take  place  before  the
spin-flip   transition  sets   in,  as   we  observed   for  high   $Z$  atoms
\citep{ivanov00}. \citet{ivanov00} investigated   the  ground  state  configurations  at
$\beta\sim  10^{-1}-10^4$   for  $Z=1-10$   atoms.  For  $Z>10$   atoms,  only
\citet{neuhauser87}  provided the  ground state  configurations and  the total
energies among the self-consistent methods.
 
Our  single-configurational   total  energies   match  very  well   with  1DHF
\citep{neuhauser87} up to $Z=5$, while  1DHF total energies become higher than
our values by at most 1\% at $Z>5$. This is probably because their grid points
are  not  sufficiently  fine for  inner  electrons  in  high $Z$  atoms.   Our
multi-configurational results are $5-10$\% lower than 1DHF at $\beta_Z \le 10$
and a few \%  lower at $\beta_Z \le 100$. Some of  their ground state energies
are  even higher  than  our multi-configurational  energy  of several  excited
states.

\citet{jones_pb86}  studied  the ground  state  energy  for  several high  $Z$
neutral atoms  by DF. Their results  are a few  \% lower than our  results and
2DHF for  the ground states  of several atoms,  while they fixed  $n=0$ (table
\ref{ground_energy2}).  Ground  state  energies  obtained  by  DF  are  highly
dependent  on  the  form   of  the  exchange-correlation  density  functionals
\citep{jones_pb85_1},  even  though DF  can  relatively  easily implement  the
exchange and electron correlation terms in the local density approximation.

Up to $Z=10$, our multi-configurational  ground state energies are higher than
2DHF results by $<$  1\% overall at $\beta_Z > 1$. Even  though 2DHF claims no
numerical  errors by  the use  of  the Richardson  extrapolation method,  this
difference could  be attributed to  the fact that  we did not include  the 2nd
order perturbation in the total energy. The errors are expected to come mainly
from the inner  electrons but to be smaller for  outer electrons. However, the
inner  electrons of high $Z$ atoms are irrelevant  to X-ray  spectroscopy since
their  binding  energy  is relatively  high  ($>$  10  keV), while  the  inner
electrons in low $Z$ atoms can  contribute to the opacities in the X-ray band.
The accuracy in  total energies within 1\% in comparison  with 2DHF results is
sufficient for the identification of  the potential atomic lines or edges. Our
energies as well  as the 0th order wavefunctions can be  improved by adding an
appropriate  local exchange  term  to a  single-orbital  Hartree equation.  In
contrast, empirical  density functionals for  the exchange term in  the Landau
regime are not known presently.

\subsection{Oscillator strengths}
\label{os}

Compared to  energy values,  oscillator strengths are  subject to  more atomic
model  dependent uncertainties.  This  is because  atomic  models are  usually
optimized for  minimizing the  total energy. The  completeness of  the orbital
wavefunctions represented by Thomus-Kuhn rules is a secondary issue. Errors in
transition  energy  values  can  affect  the  oscillator  strengths.  For  the
candidate  strong transitions, 10\%  accuracy in  the oscillator  strengths is
sufficient,  considering  that  spectral  features  can undergo  a  number  of
broadening  effects  \citep{rajagopal97}. For  the  same  reasons,  we do  not
consider transitions  other than dipole-allowed.  For hydrogen, we  compare to
the results  from MCHF  \citep{ruder} and we  included a correction by the
proton cyclotron term for  $\Delta m \ne 0$  transitions.  Consequently, the  oscillator
strengths match fairly well with their results (table \ref{h_os}).

Data   for    oscillator   strengths   at   $Z>1$   were    found   only   for
1DHF. \citet{miller90} published data  at $Z=2,6$ for bound-bound transitions,
while  \citet{ruder} obtained  the oscillator  strengths only  for helium-like
atoms.  It turns  out that tight-tight transitions match  with 1DHF very well,
while the  oscillator strengths  for tight-loose transitions  are consistently
larger than 1DHF results by $10-50$ \%. The most likely explanation is that we
did  not include  the exchange  effects in  the orbital  wavefunction.  In the
antisymmetrized  Hartree total  wavefunctions,  electrons are  closer to  each
other than  in the Hartree-Fock  total wavefunctions. As exchange  effects are
larger for  a pair of  closer electrons, \tb  electrons are more  deviant from
Hartree-Fock orbitals. This  discrepancy is small for the  inner \tb electrons
since their behavior is mainly  determined by the Coulomb attraction from the
nucleus. Also, it  is very small for  a \lb electron simply because  it is far
from  other  electrons, unless  there  are  other  \lb electrons  nearby.  For
tight-tight transitions,  a transitioning electron (usually  the outermost \tb
electron)  in  the  Hartree scheme  is  closer  to  the  nucleus than  in  the
Hartree-Fock  scheme both  in  the  initial and  final  state. This  deviation
cancels  out when  we evaluate  the oscillator  strengths by  the  length form
because a  change in the spatial  distribution of the  electron associated with
the atomic  transition is about the  same in the Hartree  and the Hartree-Fock
picture.  However,  for tight-loose transitions, the deviation  in the initial
state remains  in the  oscillator strengths and  it gives us  the consistently
larger oscillator strengths because  the Hartree scheme requires larger change
in the spatial distribution compared to the Hartree-Fock scheme.
 
A well-defined measure  of this deviation associated with  each bound electron
is  the ratio  of  the exchange  term  to the  Hartree single-orbital  energy,
i.e.   $r_{ex}=|\epsilon^{(1)}_{m\nu}/\epsilon^{(0)}_{m\nu}|$.  Therefore,  we
introduced  a  correction to  the  oscillator  strengths  for the  tight-loose
transitions by  the amount of $r_{ex}$.  The same correction is applicable
when size of atoms or ions is required to include non-ideal effects,  for
instance through the excluded-volume term \citep{potekhin99}. In contrast to hydrogen  case, we did
not include the finite nuclear  mass correction since the previous results for
comparison  assumed the  infinite nuclear  mass.  As a  result, the  corrected
oscillator  strengths  match  with  1DHF  results within  10\%  for  different
tight-loose transitions of helium and carbon at various magnetic fields (table
\ref{he_os}, \ref{c_os}).  The discrepancy  at small $\beta_Z$  is due  to the
finite occupation in  higher Landau levels, which is not  included in 1DHF but
is included  in MCPH$^3$. Although  comparisons are available only  for helium
and  carbon due to  few publications  on oscillator  strengths for  $Z>1$, the
numerical  results  in  tables   \ref{he_os}  and  \ref{c_os}  based  on  this
physically reasonable  correction, strongly support  that oscillator strengths
are calculated with 10\% accuracy for other atoms up to Fe.

%-------------------------------------------------------------------------  
\section{Future work}
Construction of an extensive atomic data base from the method presented in
this article is in progress. It consists of energies and oscillator strengths for a large number of
grid points in $B$. Based on the atomic data, equation of state and 
opacity tables will be constructed in the phase space of ($B, \rho, T, Z$) 
covering the typical parameter values for cooling neutron stars.
%-----------------------------------------------------------------------------
\acknowledgments{We are grateful to Ehud Behar for careful reading the
manuscript. This work was partially supported by NASA grant NAG5-7737.}
% ----------------------------------------------------------------------------

\begin{appendix}
\section{Evaluation of potentials and energies}
\subsection{Analytical expression of potentials and energies}
In order to avoid the complications caused by the $r^{-1}$ cusp at the origin, we
expanded the terms $r^{-1}$ and
$|\vec{r}-\vec{r}\,'|^{-1}$ appearing in the potentials  by the Bessel functions
$J_m(x)$ \citep{flowers77}. $L_m(x)$, $L^n_m(x)$ denote the Laguerre function and associated
Laguerre function respectively. Hereafter, 
$\int{dk}=\int_0^{\infty}{dk}, \int{dz}=\int_{-\infty}^{\infty}{dz},
\int{d\rho}=\int_0^{\infty}{d\rho}, \int{d\phi}=\int_0^{2\pi}{d\phi}$ .
\begin{eqnarray}
r^{-1}&=&\frac{1}{\sqrt{\rho^2+z^2}}=\int{dk}J_0(k\rho)e^{-k|z|},\\
|\vec{r}-\vec{r}\,'|^{-1}&=&\frac{1}{\sqrt{\rho^2+{\rho'}^2-2\rho\rho'\cos(\phi-\phi')+(z-z')^2}}\nonumber\\
&=&\sum_{m=-\infty}^{\infty}\int{dk}\;e^{im(\phi-\phi')}J_m(k\rho)J_m(k\rho')e^{-k|z-z'|}.
\end{eqnarray}
For the single-configurational  potentials,
\begin{eqnarray}
V_{nuc}(z)&=&-\frac{Ze^2}{\hat\rho}V_{m00}(z),\\
\nonumber\\
V_{eff}(z)&=&\frac{e^2}{\hat\rho}\sum_{m'\nu'}\int{dz'}\;D^{m00}_{m'00}(z-z')|f_{m'\nu'}(z')|^2,
\end{eqnarray}
where nuclear and direct kernels are defined as 
\begin{eqnarray}
V_{mnn'}(z)&\equiv&\int{\rho d\rho}\int{d\phi}\;\Phi^*_{n'm}(\rho,\phi)\Phi_{nm}(\rho,\phi)\;\frac{1}{r},\\
D^{mnn'}_{m'\tilde{n}\tilde{n}'}(z-z')&\equiv&\int{\rho
d\rho} \int{d\phi}\int{\rho' d\rho'}\int{d\phi'}\; \Phi^*_{\tilde
n m'}(\rho', \phi') \Phi_{\tilde n' m'}(\rho', \phi')\nonumber\\
&&\times\Phi_{nm}(\rho, \phi)\Phi^*_{n'm}(\rho, \phi)\frac{1}{|\vec{r}-\vec{r}\,'|}.
\end{eqnarray}
The matrix elements for the multi-configurational  calculation are : 
\begin{eqnarray}
\langle{nm\nu}|V_{nuc}(\vec{r})|n'm\nu\rangle&=&-\frac{Ze^2}{\hat\rho}\int{dz}\;V_{mnn'}(z)|f_{m\nu}(z)|^2,\\
\langle{nm\nu}|V_{eff}(\vec{r})|n'm\nu\rangle&=&\frac{e^2}{\hat\rho}\sum_{m'\nu'}\sum_{\tilde{n},
\tilde{n}'}c_{\tilde{n}m'\nu'}c_{\tilde{n}'m'\nu'}\int{dz}\;D^{mnn'}_{m'\tilde{n}\tilde{n}'}(z-z')\nonumber\\
& &\times |f_{m\nu}(z)|^2|f_{m'\nu'}(z')|^2.
\end{eqnarray}
1st order perturbation of single-orbital energy and total energy are given as, 
\begin{eqnarray}
\label{eq_energy1}
\epsilon^{(1)}_{m\nu}&=&-\frac{e^2}{\hat\rho}\sum_{m'\nu'}\sum_{n,
n'}\sum_{\tilde{n},
\tilde{n}'}c_{nm\nu}c_{n'm\nu}c_{\tilde{n}m'\nu'}c_{\tilde{n}'m'\nu'}\int{dz}\int{dz'}\;E^{mnn'}_{m'\tilde{n}\tilde{n}'}(z-z')\nonumber\\
& &\times f_{m\nu}(z)f_{m'\nu'}(z')f^*_{m'\nu'}(z)f^*_{m\nu}(z'),\\
\label{eq_energy2}
E^{(1)}&=&-\frac{e^2}{2\hat\rho}\sum_{m\nu}\sum_{m'\nu'}\sum_{n,
n'}\sum_{\tilde{n},
\tilde{n}'}c_{nm\nu}c_{n'm\nu}c_{\tilde{n}m'\nu'}c_{\tilde{n}'m'\nu'}\int{dz}\int{dz'}\nonumber\\
& &\Bigg(D^{mnn'}_{m'\tilde{n}\tilde{n}'}(z-z')|f_{m\nu}(z)|^2|f_{m'\nu'}(z')|^2\nonumber\\
&
&+E^{mnn'}_{m'\tilde{n}\tilde{n}'}(z-z')f_{m\nu}(z)f_{m'\nu'}(z')f^*_{m'\nu'}(z)f^*_{m\nu}(z')\Bigg), 
\end{eqnarray}
where exchange kernels are defined as 
\begin{eqnarray}
E^{mnn'}_{m'\tilde{n}\tilde{n}'}(z-z')\equiv& &\int{\rho
d\rho}\int{d\phi}\int{\rho' d\rho'}\int{d\phi'}\; \Phi^*_{n' m}(\rho', \phi') \Phi^*_{\tilde n' m'}(\rho', \phi')\nonumber\\
& &\times\frac{1}{|\vec{r}-\vec{r}\,'|}\Phi_{nm}(\rho, \phi)\Phi_{\tilde{n} m'}(\rho, \phi).
\end{eqnarray}

\subsection{Nuclear and direct kernels}
\begin{eqnarray}
V_{mnn'}(z)&=&\int{dk}\;e^{-k|z|}\zeta^m_{nn'}(k),\\
\nonumber\\
D^{mnn'}_{\tilde{m}\tilde{n}\tilde{n}'}(z-z')&=&\int{dk}\;e^{-k|z-z'|}\zeta^m_{nn'}(k)\zeta^{\tilde{m}}_{\tilde{n}\tilde{n}'}(k).
\end{eqnarray}
The zeta function is defined as,
\begin{eqnarray}
\zeta^{m}_{nn'}(k)&\equiv&\int{\rho{d\rho}}\int{d\phi}\;\Phi^*_{nm}(\rho,\phi)\Phi_{n'm}(\rho,\phi)J_0(k\rho),\nonumber\\
&=&\left[\frac{n!n'!}{2^{2m}(n+m)!(n'+m')!}\right]^{1/2}\int{d\rho}\;\rho^{2m+1}e^{-\rho^2/2}\nonumber\\
& &\times L^m_n\left(\frac{\rho^2}{2}\right)L^m_{n'}\left(\frac{\rho^2}{2}\right)J_0(k\rho).
\end{eqnarray}
$\zeta$ functions are analytically computed for some special cases.
\begin{eqnarray}
\zeta^m_{00}(k)&=&L_m\left(\frac{k^2}{2}\right)e^{-k^2/2},\\
\nonumber\\
\zeta^0_{nn'}(k)&=&(-1)^{n+n'}e^{-k^2/2}L^{n'-n}_n\left(\frac{k^2}{2}\right)L^{n-n'}_{n'}\left(\frac{k^2}{2}\right).
\end{eqnarray}
The rest of the $\zeta$ functions are computed by the following recursion
relation.
\begin{eqnarray}
\zeta^{m}_{nn'}(k)&=&\sum_{n''=0}^{n'}\left[\frac{n'!(n''+m-1)!}{n''!(n'+m)!}\right]^{1/2}\left[\sqrt{n+m}\;\zeta^{m-1}_{nn''}(k)-\sqrt{n+1}\;\zeta^{m-1}_{n+1n''}(k)\right],\\
\nonumber\\
\zeta^{m}_{n'n}(k)&=&\zeta^{m}_{nn'}(k).
\end{eqnarray}

\subsection{Exchange kernels}
\begin{equation}
E^{mnn'}_{m'\tilde{n}\tilde{n}'}(z-z')=\int{dk} \;
e^{-k|z-z'|}\eta^{m'\tilde{n}}_{mn}(k)\eta^{m'\tilde{n}'}_{m n'}(k).
\end{equation}
The eta function is defined as, 
\begin{eqnarray}
\eta^{m'n'}_{mn}(k)&\equiv&\int{\rho{d\rho}}\int{d\phi}\;\Phi^*_{nm}(\rho,\phi)\Phi_{n'm'}(\rho,\phi)J_{m'-m}(k\rho)\nonumber\\
&=&\left[\frac{n!n'!}{2^{m+m'}(n+m)!(n'+m')!}\right]^{1/2}\int{d\rho}\;\rho^{m+m'+1}e^{-\rho^2/2}\nonumber\\
& &\times L^m_n\left(\frac{\rho^2}{2}\right)L^{m'}_{n'}\left(\frac{\rho^2}{2}\right)J_{m'-m}(k\rho).
\end{eqnarray}
$\eta$ functions are analytically computed for some special cases.
\begin{equation}
\eta^{0n'}_{mn}(k)=(-1)^{n+n'}\left[\frac{n!}{2^m(n+m)!}\right]^{1/2}k^me^{-k^2/2}L^{n'-n}_n\left(\frac{k^2}{2}\right)L^{m+n-n'}_{n'}\left(\frac{k^2}{2}\right).
\end{equation}
The rest of the $\eta$ functions are computed by the following recursion
relation.
\begin{eqnarray}
\eta^{m'n'}_{mn}(k)&=&\sum_{n''=0}^{n'}\left[\frac{n'!(n''+m-1)!}{n''!(n'+m)!}\right]^{1/2}[\sqrt{2(m'-m-1)^2}\;\frac{\eta^{m'-1n''}_{mn}(k)}{k}\nonumber\\
& &-\sqrt{n+m+1}\;\eta^{m'-1n''}_{m+1n}(k)+\sqrt{2n}\;\eta^{m'-1n''}_{m+1n-1}(k)],\\
\nonumber\\
\eta^{mn'}_{m'n}(k)&=&\eta^{m'n}_{mn'}(k)=\left[\frac{(n+m')!(n'+m)!}{(n+m)!(n'+m')!}\right]^{1/2}\eta^{m'n'}_{mn}(k).
\end{eqnarray}
\end{appendix}

%\bibliography{list}

\begin{thebibliography}{88}
\expandafter\ifx\csname natexlab\endcsname\relax\def\natexlab#1{#1}\fi

\bibitem[{{Abrahams} \& {Shapiro}(1991)}]{abrahams}
{Abrahams}, A.~M. \& {Shapiro}, S.~L. 1991, \apj, 382, 233

\bibitem[{{Alcock} \& {Illarionov}(1980)}]{alcock80}
{Alcock}, C. \& {Illarionov}, A. 1980, \apj, 235, 534

\bibitem[{{Baye} \& {Vincke}(1990)}]{baye90}
{Baye}, D. \& {Vincke}, M. 1990, \pra, 42, 391

\bibitem[{{Becken} \& {Schmelcher}(2000)}]{becken00}
{Becken}, W. \& {Schmelcher}, P. 2000, J. Phys. B, 33, 545

\bibitem[{{Becken} \& {Schmelcher}(2001)}]{becken01}
---. 2001, \pra, 63, 0a4+

\bibitem[{{Becken} {et~al.}(1999){Becken}, {Schmelcher}, \&
  {Diakonos}}]{becken99}
{Becken}, W., {Schmelcher}, P., \& {Diakonos}, F.~K. 1999, J. Phys. B, 32, 1557

\bibitem[{{Becker} \& {Tr{\"u}mper}(1997)}]{becker97}
{Becker}, W. \& {Tr{\"u}mper}, J. 1997, \aap, 326, 682

\bibitem[{{Bezchastnov} {et~al.}(1998){Bezchastnov}, {Pavlov}, \&
  {Ventura}}]{bezchastnov98}
{Bezchastnov}, V.~G., {Pavlov}, G.~G., \& {Ventura}, J. 1998, \pra, 58, 180

\bibitem[{{Camilo} {et~al.}(1994){Camilo}, {Thorsett}, \&
  {Kulkarni}}]{camilo94}
{Camilo}, F., {Thorsett}, S.~E., \& {Kulkarni}, S.~R. 1994, \apjl, 421, L15

\bibitem[{{Chen} {et~al.}(1974){Chen}, {Ruderman}, \& {Sutherland}}]{chenh74}
{Chen}, H., {Ruderman}, M.~A., \& {Sutherland}, P.~G. 1974, \apj, 191, 473

\bibitem[{{Chen} \& {Goldman}(1992)}]{chenz92}
{Chen}, Z. \& {Goldman}, S.~P. 1992, \pra, 45, 1722

\bibitem[{{Chevalier}(1996)}]{chevalier96}
{Chevalier}, R.~A. 1996, \apj, 459, 322

\bibitem[{{Demeur} {et~al.}(1994){Demeur}, {Heenen}, \& {Godefroid}}]{demeur94}
{Demeur}, M., {Heenen}, P.~., \& {Godefroid}, M. 1994, \pra, 49, 176

\bibitem[{{Flowers} {et~al.}(1977){Flowers}, {Ruderman}, {Lee}, {Sutherland},
  {Hillebrandt}, \& {Mueller}}]{flowers77}
{Flowers}, E.~G., {Ruderman}, M.~A., {Lee}, J.~., {Sutherland}, P.~G.,
  {Hillebrandt}, W., \& {Mueller}, E. 1977, \apj, 215, 291

\bibitem[{{Forster} {et~al.}(1984){Forster}, {Strupat}, {Rosner}, {Wunner},
  {Ruder}, \& {Herold}}]{forster84}
{Forster}, H., {Strupat}, W., {Rosner}, W., {Wunner}, G., {Ruder}, H., \&
  {Herold}, H. 1984, J. Phys. B, 17, 1301

\bibitem[{{Glasser}(1975)}]{glasser75_2}
{Glasser}, M.~L. 1975, \apj, 199, 206

\bibitem[{{Glasser} \& {Kaplan}(1975)}]{glasser75_1}
{Glasser}, M.~L. \& {Kaplan}, J.~I. 1975, \apj, 199, 208

\bibitem[{{Hewish} {et~al.}(1968){Hewish}, {Bell}, {Pilkington}, {Scott}, \&
  {Collins}}]{hewish68}
{Hewish}, A.~S., {Bell}, A.~J., {Pilkington}, J. D.~H., {Scott}, P.~F., \&
  {Collins}, R.~A. 1968, \nat, 217, 709

\bibitem[{{Ho} \& {Lai}(2001)}]{ho01}
{Ho}, W. C.~G. \& {Lai}, D. 2001, preprint (astro-ph/0104199)

\bibitem[{{Holas} \& {March}(1997)}]{holas97}
{Holas}, A. \& {March}, N.~H. 1997, \pra, 56, 4595

\bibitem[{{Hylleraas}(1930)}]{hylleras30}
{Hylleraas}, E.~A. 1930, Z. Physik, 65, 209

\bibitem[{{Ivanov}(1994)}]{ivanov94}
{Ivanov}, M.~V. 1994, J. Phys. B, 27, 4513

\bibitem[{{Ivanov} \& {Schmelcher}(1998)}]{ivanov98}
{Ivanov}, M.~V. \& {Schmelcher}, P. 1998, \pra, 57, 3793

\bibitem[{{Ivanov} \& {Schmelcher}(1999)}]{ivanov99}
---. 1999, \pra, 60, 3558

\bibitem[{{Ivanov} \& {Schmelcher}(2000)}]{ivanov00}
---. 2000, \pra, 61, 022505

\bibitem[{{Johnsen} \& {Yngvason}(1996)}]{johnsen}
{Johnsen}, K. \& {Yngvason}, J. 1996, \pra, 54, 1936

\bibitem[{{Jones} {et~al.}(1997){Jones}, {Ortiz}, \& {Ceperley}}]{jones_md97}
{Jones}, M.~D., {Ortiz}, G., \& {Ceperley}, D.~M. 1997, \pre, 55, 6202

\bibitem[{{Jones} {et~al.}(1999{\natexlab{a}}){Jones}, {Ortiz}, \&
  {Ceperley}}]{jones_md99_2}
---. 1999{\natexlab{a}}, \aap, 343, L91

\bibitem[{{Jones} {et~al.}(1999{\natexlab{b}}){Jones}, {Ortiz}, \&
  {Ceperley}}]{jones_md99_1}
---. 1999{\natexlab{b}}, \pra, 59, 2875

\bibitem[{{Jones}(1985{\natexlab{a}})}]{jones_pb85_1}
{Jones}, P.~B. 1985{\natexlab{a}}, \prl, 55, 1338

\bibitem[{{Jones}(1985{\natexlab{b}})}]{jones_pb85_2}
---. 1985{\natexlab{b}}, \mnras, 216, 503

\bibitem[{{Jones}(1986)}]{jones_pb86}
---. 1986, \mnras, 218, 477

\bibitem[{{Jordan} {et~al.}(1998){Jordan}, {Schmelcher}, {Becken}, \&
  {Schweizer}}]{jordan98}
{Jordan}, S., {Schmelcher}, P., {Becken}, W., \& {Schweizer}, W. 1998, \aap,
  336, L33

\bibitem[{{Koonin}(1986)}]{koonin86}
{Koonin}, S.~E. 1986, Computational Physics (Benjamin/Cummings, Menlo Park,
  California)

\bibitem[{{Kopidakis} {et~al.}(1996){Kopidakis}, {Ventura}, \&
  {Herold}}]{kopidakis96}
{Kopidakis}, N., {Ventura}, J., \& {Herold}, H. 1996, \aap, 308, 747

\bibitem[{{K{\"o}ssl} {et~al.}(1988){K{\"o}ssl}, {Wolff}, {M{\"u}ller}, \&
  {Hillebrandt}}]{koessl88}
{K{\"o}ssl}, D., {Wolff}, R.~G., {M{\"u}ller}, E., \& {Hillebrandt}, W. 1988,
  \aap, 205, 347

\bibitem[{{Lai}(2000)}]{lai00}
{Lai}, D. 2000, preprint (astro-ph/0009333)

\bibitem[{{Lai} \& {Salpeter}(1995)}]{lai95}
{Lai}, D. \& {Salpeter}, E.~E. 1995, \pra, 52, 2611

\bibitem[{{Lai} \& {Salpeter}(1996)}]{lai96}
---. 1996, \pra, 53, 152

\bibitem[{{Lai} \& {Salpeter}(1997)}]{lai97}
---. 1997, \apj, 491, 270

\bibitem[{{Lai} {et~al.}(1992){Lai}, {Salpeter}, \& {Shapiro}}]{lai92}
{Lai}, D., {Salpeter}, E.~E., \& {Shapiro}, S.~L. 1992, \pra, 45, 4832

\bibitem[{{Landau} \& {Lifshitz}(1965)}]{landau65}
{Landau}, L.~D. \& {Lifshitz}, E.~M. 1965, Quantum mechanics (Course of
  theoretical physics, Oxford: Pergamon Press, 1965)

\bibitem[{{Lieb} {et~al.}(1992){Lieb}, {Solovej}, \& {Yngvason}}]{lieb}
{Lieb}, E.~H., {Solovej}, J.~P., \& {Yngvason}, J. 1992, \prl, 69, 749

\bibitem[{{Lindgren} \& {Virtamo}(1979)}]{lindgren79}
{Lindgren}, K. A.~U. \& {Virtamo}, J.~T. 1979, J. Phys. B, 12, 3465

\bibitem[{{Meszaros}(1992)}]{meszaros92}
{Meszaros}, P. 1992, High-energy radiation from magnetized neutron stars
  (Theoretical Astrophysics, Chicago: University of Chicago Press, |c1992)

\bibitem[{{Miller}(1990)}]{miller90}
{Miller}, M.~C. 1990, PhD thesis, California Inst.\ of Tech., Pasadena.

\bibitem[{{Miller}(1992)}]{miller92}
---. 1992, \mnras, 255, 129

\bibitem[{{Miller} \& {Neuhauser}(1991)}]{miller91}
{Miller}, M.~C. \& {Neuhauser}, D. 1991, \mnras, 253, 107

\bibitem[{{M{\"u}ller}(1984)}]{mueller84}
{M{\"u}ller}, E. 1984, \aap, 130, 415

\bibitem[{{Neuhauser} {et~al.}(1987){Neuhauser}, {Koonin}, \&
  {Langanke}}]{neuhauser87}
{Neuhauser}, D., {Koonin}, S.~E., \& {Langanke}, K. 1987, \pra, 36, 4163

\bibitem[{{Neuhauser} {et~al.}(1986){Neuhauser}, {Langanke}, \&
  {Koonin}}]{neuhauser86}
{Neuhauser}, D., {Langanke}, K., \& {Koonin}, S.~E. 1986, \pra, 33, 2084

\bibitem[{{Ortiz} {et~al.}(1995){Ortiz}, {Jones}, \& {Ceperley}}]{ortiz95}
{Ortiz}, G., {Jones}, M.~D., \& {Ceperley}, D.~M. 1995, \pra, 52, 3405

\bibitem[{{Ostriker} \& {Gunn}(1969)}]{ostriker69}
{Ostriker}, J.~P. \& {Gunn}, J.~E. 1969, \nat, 221, 454

\bibitem[{{{\"O}zel}(2001)}]{ozel01}
{{\"O}zel}, F. 2001, preprint (astro-ph/0103227)

\bibitem[{{Paerels} {et~al.}(2001){Paerels}, {Mori}, {Motch}, {Haberl},
  {Zavlin}, {Zane}, {Ramsay}, {Cropper}, \& {Brinkman}}]{paerels01}
{Paerels}, F., {Mori}, K., {Motch}, C., {Haberl}, F., {Zavlin}, V.~E., {Zane},
  S., {Ramsay}, G., {Cropper}, M., \& {Brinkman}, B. 2001, \aap, 365, L298

\bibitem[{{Page}(1998)}]{page98}
{Page}, D. 1998, in NATO ASIC Proc. 515: The Many Faces of Neutron Stars., 539

\bibitem[{{Pavlov} \& {Meszaros}(1993)}]{pavlov93}
{Pavlov}, G.~G. \& {Meszaros}, P. 1993, \apj, 416, 752

\bibitem[{{Pavlov} {et~al.}(1995){Pavlov}, {Shibanov}, {Zavlin}, \&
  {Meyer}}]{pavlov95_1}
{Pavlov}, G.~G., {Shibanov}, Y.~A., {Zavlin}, V.~E., \& {Meyer}, R.~D. 1995, in
  The Lives of the Neutron Stars. Proceedings of the NATO Advanced Study
  Institute on the Lives of the Neutron Stars, Publisher, Kluwer Academic,
  Dordrecht, The Netherlands, Boston, Massachusetts, 1995. p.71

\bibitem[{{Pavlov} {et~al.}(2001){Pavlov}, {Zavlin}, {Sanwal}, {Burwitz}, \&
  {Garmire}}]{pavlov01}
{Pavlov}, G.~G., {Zavlin}, V.~E., {Sanwal}, D., {Burwitz}, V., \& {Garmire}, G.
  2001, \apjl, 552, L129

\bibitem[{{Pavlov-Verevkin} \& {Zhilinskii}(1980)}]{pavlov-verevkin80-1}
{Pavlov-Verevkin}, V.~B. \& {Zhilinskii}, B. 1980, Phys. Lett. A, 78, 244

\bibitem[{{Pons} {et~al.}(2000){Pons}, {Walter}, {Lattimer}, \&
  {Prakash}}]{pons00}
{Pons}, J.~A., {Walter}, F.~W., {Lattimer}, J.~M., \& {Prakash}, M. 2000,
  AAS/High Energy Astrophysics Division, 32, 3314

\bibitem[{{Potekhin}(1994)}]{potekhin94}
{Potekhin}, A.~Y. 1994, J. Phys. B, 27, 1073

\bibitem[{{Potekhin}(1998)}]{potekhin98-2}
---. 1998, J. Phys. B, 31, 49

\bibitem[{{Potekhin} {et~al.}(1999){Potekhin}, {Chabrier}, \&
  {Shibanov}}]{potekhin99}
{Potekhin}, A.~Y., {Chabrier}, G., \& {Shibanov}, Y.~A. 1999, \pre, 60, 2193

\bibitem[{{Rajagopal} \& {Romani}(1996)}]{rajagopal96}
{Rajagopal}, M. \& {Romani}, R.~W. 1996, \apj, 461, 327

\bibitem[{{Rajagopal} {et~al.}(1997){Rajagopal}, {Romani}, \&
  {Miller}}]{rajagopal97}
{Rajagopal}, M., {Romani}, R.~W., \& {Miller}, M.~C. 1997, \apj, 479, 347

\bibitem[{{Relovsky} \& {Ruder}(1996)}]{relovsky}
{Relovsky}, B.~M. \& {Ruder}, H. 1996, \pra, 53, 4068

\bibitem[{{R{\"o}gnvaldsson} {et~al.}(1993){R{\"o}gnvaldsson}, {Fushiki},
  {Gudmundsson}, {Pethick}, \& {Yngvason}}]{roegnvaldsson}
{R{\"o}gnvaldsson}, O.~E., {Fushiki}, I., {Gudmundsson}, E.~H., {Pethick},
  C.~J., \& {Yngvason}, J. 1993, \apj, 416, 276

\bibitem[{{Romani}(1987)}]{romani87}
{Romani}, R.~W. 1987, \apj, 313, 718

\bibitem[{{R{\"o}sner} {et~al.}(1984){R{\"o}sner}, {Wunner}, {Herold}, \&
  {Ruder}}]{roesner84}
{R{\"o}sner}, W., {Wunner}, G., {Herold}, H., \& {Ruder}, H. 1984, J. Phys. B,
  17, 29

\bibitem[{{Ruder} {et~al.}(1994){Ruder}, {Wunner}, {Herold}, \&
  {Geyer}}]{ruder}
{Ruder}, H., {Wunner}, G., {Herold}, H., \& {Geyer}, F. 1994, Atoms in Strong
  Magnetic Fields. Quantum Mechanical Treatment and Applications in
  Astrophysics and Quantum Chaos (X, 309 pp.\ 93 figs..\ Springer-Verlag Berlin
  Heidelberg New York.)

\bibitem[{{Ruderman}(1971)}]{ruderman71}
{Ruderman}, M.~A. 1971, \prl, 27, 1306

\bibitem[{{Salpeter}(1998)}]{salpeter98}
{Salpeter}, E.~E. 1998, Journal de Physique, 10, 11285

\bibitem[{{Schmelcher}(1995)}]{schmelcher95}
{Schmelcher}, P. 1995, \pra, 52, 130

\bibitem[{{Schmelcher} {et~al.}(1999){Schmelcher}, {Ivanov}, \&
  {Becken}}]{schmelcher99}
{Schmelcher}, P., {Ivanov}, M.~V., \& {Becken}, W. 1999, \pra, 59, 3424

\bibitem[{{Scrinzi}(1998)}]{scrinzi}
{Scrinzi}, A. 1998, \pra, 58, 3879

\bibitem[{{Skjervold} \& {Ostgaard}(1984)}]{skjervold84}
{Skjervold}, J.~E. \& {Ostgaard}, E. 1984, \physscr, 29, 543

\bibitem[{{Steinberg} {et~al.}(1998){Steinberg}, {Ortner}, \&
  {Ebeling}}]{steinberg98}
{Steinberg}, M., {Ortner}, J., \& {Ebeling}, W. 1998, \pre, 58, 3806

\bibitem[{{Thorolfsson} {et~al.}(1998){Thorolfsson}, {R{\"o}gnvaldsson},
  {Yngvason}, \& {Gudmundsson}}]{thorolfsson}
{Thorolfsson}, A., {R{\"o}gnvaldsson}, O.~E., {Yngvason}, J., \& {Gudmundsson},
  E.~H. 1998, \apj, 502, 847

\bibitem[{{Thurner} {et~al.}(1993){Thurner}, {Korbel}, {Braun}, {Herold},
  {Ruder}, \& {Wunner}}]{thurner}
{Thurner}, G., {Korbel}, H., {Braun}, M., {Herold}, H., {Ruder}, H., \&
  {Wunner}, G. 1993, J. Phys. B, 26, 4719

\bibitem[{{Tr{\"u}mper} {et~al.}(1978){Tr{\"u}mper}, {Pietsch}, {Reppin},
  {Voges}, {Staubert}, \& {Kendziorra}}]{trumper}
{Tr{\"u}mper}, J., {Pietsch}, W., {Reppin}, C., {Voges}, W., {Staubert}, R., \&
  {Kendziorra}, E. 1978, \apjl, 219, L105

\bibitem[{{Tsuruta}(1995)}]{tsuruta95}
{Tsuruta}, S. 1995, in The Lives of the Neutron Stars. Proceedings of the NATO
  Advanced Study Institute on the Lives of the Neutron Stars, Publisher, Kluwer
  Academic, Dordrecht, The Netherlands, Boston, Massachusetts, 1995. p.133

\bibitem[{{Vincke} {et~al.}(1992){Vincke}, {LeDourneuf}, \& {Baye}}]{vincke92}
{Vincke}, M., {LeDourneuf}, M., \& {Baye}, D. 1992, Journal of Physics B Atomic
  Molecular Physics, 25, 2787

\bibitem[{{Wheaton} {et~al.}(1979){Wheaton}, {Doty}, {Primini}, {Cooke},
  {Dobson}, {Goldman}, {Hecht}, {Howe}, {Hoffman}, \& {Scheepmaker}}]{wheaton}
{Wheaton}, W.~A., {Doty}, J.~P., {Primini}, F.~A., {Cooke}, B.~A., {Dobson},
  C.~A., {Goldman}, A., {Hecht}, M., {Howe}, S.~K., {Hoffman}, J.~A., \&
  {Scheepmaker}, A. 1979, \nat, 282, 240

\bibitem[{{Woosley} \& {Weaver}(1986)}]{woosley86}
{Woosley}, S.~E. \& {Weaver}, T.~A. 1986, \araa, 24, 205

\bibitem[{{Yakovlev} {et~al.}(1999){Yakovlev}, {Levenfish}, \&
  {Shibanov}}]{yakovlev99}
{Yakovlev}, D.~G., {Levenfish}, K.~P., \& {Shibanov}, Y.~A. 1999,
  Physics-Uspekhi, 42, 737

\bibitem[{{Zane} {et~al.}(2000){Zane}, {Turolla}, \& {Treves}}]{zane00}
{Zane}, S., {Turolla}, R., \& {Treves}, A. 2000, \apj, 537, 387

\bibitem[{{Zavlin} \& {Pavlov}(1998)}]{zavlin98}
{Zavlin}, V.~E. \& {Pavlov}, G.~G. 1998, \aap, 329, 583

\bibitem[{{Zavlin} {et~al.}(1996){Zavlin}, {Pavlov}, \& {Shibanov}}]{zavlin96}
{Zavlin}, V.~E., {Pavlov}, G.~G., \& {Shibanov}, Y.~A. 1996, \aap, 315, 141

\end{thebibliography}
%\bibliographystyle{apj}

% ==========================================================================
% Figures
% ==========================================================================

% =========================================================================
% Tables
% =========================================================================

\begin{deluxetable}{ccccccccc}
\tabletypesize{\scriptsize}
\tablewidth{0pt}
\tablecaption{Energies of \tb states for hydrogen. \label{h_ground_energy}}
\tablehead{
\colhead{$(m, \nu)$} & \multicolumn{2}{c}{(0,0)} & \multicolumn{2}{c}{(1,0)} &
 \multicolumn{2}{c}{(2,0)} & \multicolumn{2}{c}{(3,0)}\\
\\
$\beta$ & \colhead{Present work} & \colhead{Ruder\tablenotemark{a}} &
\colhead{Present work} & \colhead{Ruder} & \colhead{Present work} & \colhead{Ruder} &
\colhead{Present work} & \colhead{Ruder}
}
\startdata
1 & 28.10 & 27.82\tablenotemark{b} & 16.33 & 16.32\tablenotemark{b} & 12.81 & 12.82\tablenotemark{b}  & 10.84 &10.97\tablenotemark{b} \\
2 & 34.53 &34.85\tablenotemark{b}  & 21.38 &21.44\tablenotemark{b}&   16.84 &17.06\tablenotemark{b}  & 14.57 & 14.69\\
5 & 47.00 &47.56\tablenotemark{b}&   30.49 &30.62\tablenotemark{b}&   24.52 &24.72  & 21.33 &21.44\\
10 & 59.61 &60.26\tablenotemark{b}  & 39.49 &39.88&   32.31 &32.48  & 28.22 &28.32\\
20 & 75.50 &76.22\tablenotemark{b}  & 51.26 &51.60&   42.24 &42.39 & 37.06 &37.14  \\
50 & 102.41 &103.11  & 71.38 &71.70 & 59.42 &59.54  & 52.44 &52.52  \\
100 & 127.93 &128.64  & 90.74 &91.08  & 76.13 &76.25  & 67.52 &67.59 \\
200 & 158.50 &159.24  & 114.45 &114.70 & 96.66 &96.77  & 86.14 &86.20  \\
500 & 207.47 &208.52  & 153.17 &153.43  & 130.62 &130.76  & 117.15 &117.24 \\
1000 & 252.30 &253.21  & 188.72 &189.18  & 162.17 &162.42  & 146.17 &146.34 \\
\enddata
\tablenotetext{a}{Multi-configurational Hartree-Fock method \citep{ruder}}
\tablenotetext{b}{Energies are calculated by the spherical expansion of wave functions.}
\tablecomments{All energies are in eV. $\Delta \tilde z=2 \times 10^{-2}$.}
\end{deluxetable}

\begin{deluxetable}{ccccccccccc}
\tablewidth{0pt}
\tabletypesize{\scriptsize}
\tablecaption{Energies of \lb states for hydrogen. \label{h_excite_energy}}
\tablehead{
\colhead{$(m,\nu)$} & \multicolumn{2}{c}{(0,1)} & \multicolumn{2}{c}{(0,2)} & \multicolumn{2}{c}{(1,1)} & \multicolumn{2}{c}{(1,2)}  \\
\\
$\beta$ & \colhead{Present work} & \colhead{Ruder\tablenotemark{a}} & \colhead{Present work} & \colhead{Ruder} & \colhead{Present work} & \colhead{Ruder} & \colhead{Present work} & \colhead{Ruder}
}
\startdata
1 & 	8.051	& 8.101\tablenotemark{b} & 4.584 & 4.733\tablenotemark{b} &
6.653 & 6.673	& 3.832	& 3.876 \\ 
2 & 	9.096	& 9.135\tablenotemark{b} & 5.053 & 5.138\tablenotemark{b} & 7.759 & 7.778  & 4.314& 4.346 \\
5  & 	10.38 	& 10.41 & 5.650 & 5.684 & 9.221 & 9.224  & 4.940 & 4.961 \\
10 & 	11.24 	& 11.24 & 6.076 & 6.090  & 10.23 & 10.23 & 5.396 & 5.411 \\
20 & 	11.94 	& 11.94 & 6.480 & 6.481  & 11.10 & 11.12 & 5.834 & 5.845 \\
50 & 	12.62 	& 12.62 & 6.956 & 6.971  & 12.05 & 12.05 & 6.381 & 6.388\\
100 & 	12.96 	& 12.96 & 7.310	& 7.319  & 12.55 & 12.56 & 6.767 & 6.775\\
200 & 	13.20 	& 13.20 & 7.632	& 7.647  & 12.94 & 12.94 & 7.136 & 7.140 \\
500 & 	13.39 	& 13.40 & 8.040	& 8.051  & 13.24 & 13.25 & 7.590 & 7.590 \\
1000 & 	13.49 	& 13.49 & 8.329	& 8.333  & 13.39 & 13.39 & 7.907 & 7.906 \\
\enddata
\tablenotetext{a}{Multi-configurational Hartree-Fock method \citep{ruder}}
\tablenotetext{b}{Energies are calculated by the spherical expansion of wave functions.}
\tablecomments{All energies are in eV. $\Delta z=0.2$.}
\end{deluxetable}

\begin{deluxetable}{cccc}
\tablecaption{Ground state energies for helium. \label{he_ground_energy}}
\tablewidth{0pt}
\tablehead{
\colhead{$\beta_z$} & \colhead{Present work} & \colhead{Ruder\tablenotemark{a}} &
\colhead{Jones\tablenotemark{b}}
}
\startdata
1 	& 146.06 (134.36) 	& 134.29 	& 146.94  \\
2 	& 184.73 (176.34) 	& 176.30 	& 186.85  \\
5 	& 255.39 (249.71) 	& 249.80 	& 258.19  \\
10 	& 326.05 (322.31) 	& 322.22 	& 329.27  \\
20 	& 414.92 (412.19) 	& 412.26 	& 418.22  \\
50 	& 564.50 (562.74) 	& 563.61 	& 568.39  \\
100 	& 707.15 (705.91) 	& 706.99 	& 710.95  \\
200 	& 880.00 (879.14) 	& 879.04 	& \nodata \\
500 	& 1158.49 (1158.49) 	& 1156.63  	& \nodata \\
1000 	& 1410.90 (1410.90) 	& 1409.13  	& \nodata \\
\enddata
\tablenotetext{a}{1DHF \citep{ruder}}
\tablenotetext{b}{UHF \citep{jones_md99_1}}
\tablecomments{All energies are in eV. The ground state configuration is $(m,
\nu) = (0,0), (1,0)$. $\beta_Z=1$
for helium corresponds to $1.88\times10^{10}$ G. Bracketed values are results from our single-configurational 
calculation. $\Delta \tilde z=4\times 10^{-2}$. $N_{iter}=3$.}
\end{deluxetable}

\begin{deluxetable}{cccccc}
\tablewidth{0pt}
\tablecaption{Energies of excited states for helium at $\beta_Z=50$. \label{he_excite_energy}}
\tablehead{
\colhead{$(m,\nu)$} & \colhead{Present work} & \colhead{Miller\tablenotemark{a}} &
\colhead{Ruder\tablenotemark{b}} & \colhead{Jones\tablenotemark{c}}}
\startdata
(0,1) 	& 423.29 	& 413.7 	& 421.71 	& 426.39 \\
(0,2) 	& 418.34 	& 408.8 	& 416.17 	& 420.88 \\
(1,1) 	& 421.72 	& 412.2 	& 419.68 	& 424.38 \\
(1,2) 	& 416.61 	& 407.8 	& 415.25	& 419.95  \\
(2,0) 	& 521.73 	& \nodata 	& 521.37 	& 525.97 \\
(2,1) 	& 421.45  	& \nodata 	& 419.49 	& 424.18 \\
(2,2) 	& 416.26  	&  \nodata	& 414.83 	& 419.52 \\
\enddata
\tablenotetext{a}{1DHF \citep{miller91}} \tablenotetext{b}{1DHF \citep{ruder}}
\tablenotetext{c}{UHF \citep{jones_md99_1}}
\tablecomments{All energies are in eV. One electron is in
$(m,\nu)=(0,0)$. $\Delta \tilde z=4\times 10^{-2}$. $N_{iter}=3$.}
\end{deluxetable}

\begin{deluxetable}{ccccccc}
\tablewidth{0pt}
\tabletypesize{\small}
\tablecaption{Ground state energies of neutral atoms through $Z = 1-14$. \label{ground_energy1}}
\tablehead{
 & \multicolumn{3}{c}{$B_{12}=0.1$} & \multicolumn{3}{c}{$B_{12}=0.5$} \\
\\
$Z$ & \colhead{Present work} & \colhead{Ivanov\tablenotemark{a}} & \colhead{Neuhauser\tablenotemark{b}} & \colhead{Present work} & \colhead{Ivanov} & \colhead{Neuhauser}
}
\startdata
1 & 0.0770 (0.0762) & 0.07781 & 0.0761 & 0.1301 (0.1298) &  0.13114 & 0.130 \\

2 & 0.2613 (0.2557) &  0.26387 & 0.255 & 0.4567 (0.4543) & 0.46063 & 0.454  \\

3 & 0.5337 (0.5160) &  0.54042 & 0.516  & 0.9526 (0.9449) & 0.96180 & 0.944 \\

4 & 0.8897 (0.8468) & 0.89833 & 0.846 & 1.600 (1.580) & 1.61624 & 1.580 \\

5 & 1.327 (1.238) & 1.33229 & 1.238 & 2.390 (2.349) & 2.41101 & 2.347 \\

6 & 1.854 (1.687) & 1.83895 & 1.678 & 3.308 (3.237) & 3.33639 & 3.22 \\

7 & (2.186) & 2.41607 & 2.17 & 4.353 (4.239) & 4.38483 & 4.22 \\

8 & (2.754 [1])  & 3.08253 [1] & 2.71 & 5.517 (5.338)  & 5.55032 & 5.32 \\

9 & (3.377 [1])  & 3.82966 [1] & 3.36 & 6.803 (6.544) & 6.82794 & 6.51 \\

10 & (4.049 [1]) & 4.65087 [1] & \nodata &  8.198 (7.834) & 8.21365 & 7.819 \\

11 & \nodata 	&  \nodata& \nodata &  9.718 (9.224) & \nodata & 9.197 \\

12 & \nodata  & \nodata & \nodata & 11.410 [1] (10.750 [1])& \nodata & 10.72 [1] \\

13 & \nodata 	& \nodata  & \nodata	& 13.251 [1] (12.364 [1]) & \nodata & 12.32 [1]  \\

14 & \nodata 	& \nodata 	& \nodata  & 15.246 [2] (14.062 [2]) & \nodata & 14.00 [1] \\
\enddata
\tablenotetext{a}{2DHF \citep{ivanov00}} 
\tablenotetext{b}{1DHF \citep{neuhauser86}}
\tablecomments{All energies are in keV. The values in parentheses are
single-configurational  energy values. The number of $\nu=1$ orbitals are in
square brackets (No brackets indicate all the electrons are in $\nu=0$
orbital). $\Delta \tilde z=(2-5)\times10^{-2}$. $N_{iter}=5$.}
\end{deluxetable}

\begin{deluxetable}{ccccccccc}
\tablewidth{0pt}
\tabletypesize{\scriptsize}
\tablecaption{Ground state energies of neutral atoms through $Z = 1-26$. \label{ground_energy2}}
\tablehead{
 & \multicolumn{4}{c}{$B_{12}=1.0$} & \multicolumn{4}{c}{$B_{12}=5.0$}\\
\\
$Z$ & \colhead{Present work} &\colhead{Ivanov\tablenotemark{a}} & \colhead{Neuhauser\tablenotemark{b}} & \colhead{Jones\tablenotemark{c}} & \colhead{Present work}& \colhead{Ivanov} & \colhead{Neuhauser} & \colhead{Jones}
}
\startdata
1 & 0.1614 & 0.16222 & 0.161 & \nodata & 0.2558 & 0.25750 & 0.255 & \nodata\\

2 & 0.5766 & 0.57999 & 0.574 &\nodata & 0.9574 & 0.96191 & 0.958 & 1.040 \\

3 & 1.214 & 1.22443 & 1.209 &\nodata & 2.078 & 2.08931 &  2.076 & \nodata \\

4 & 2.056 & 2.07309 & 2.042&\nodata & 3.586 & 3.61033 & 3.584& \nodata \\

5 & 3.085 & 3.10924 & 3.054&\nodata & 5.476 & 5.49950 & 5.456 & \nodata\\

6 & 4.288 & 4.31991 & 4.20& \nodata& 7.695 & 7.73528 & 7.60 &8.03 \\

7 & 5.657 & 5.69465 & 5.54&\nodata & 10.231 & 10.29919 & 10.20 & \nodata\\

8 & 7.176 & 7.22492 & 7.02&\nodata & 13.099 & 13.17543 & 13.00 & \nodata\\

9 & 8.845 & 8.90360 & 8.63&\nodata & 16.264  & 16.34997 & 16.10 & \nodata\\

10 & 10.664 & 10.72452 & 10.39 & 10.70 & 19.702 & 19.81072 & 19.57 & 20.24\\

11 & 12.625 & \nodata & 12.25& \nodata& 23.406 & \nodata & 24.64 & \nodata\\

12 & 14.745 & \nodata & 14.23&\nodata & 27.436 & \nodata & 27.17 & \nodata \\

13 & 16.973 [1] & \nodata & 16.34 [1]& \nodata& 31.675 & \nodata & 31.35 & \nodata \\

14 & 19.408 [1] & \nodata & 18.60 [1] & 19.09 &  36.154 & \nodata & 35.74 & 36.76 \\

15 & 21.987 [1] & \nodata & 20.95 [1]&\nodata  & 40.915 & \nodata & 40.35 & \nodata \\

16 & 24.718 [2] & \nodata & 23.43 [2]& \nodata & 45.881 & \nodata & 45.22 & \nodata \\

17 & 27.618 [2] & \nodata & 26.07 [2]& \nodata & 51.067 & \nodata & 50.30 & \nodata \\

18 & 30.766 [2]  & \nodata & 28.82 [2]& \nodata & 56.530 & \nodata & 55.95 & \nodata \\

19 & 34.036 [2] & \nodata & \nodata &\nodata& 62.181 & \nodata & \nodata & \nodata\\

20 & 37.500 [3] & \nodata & \nodata & 35.48 &  68.031 & \nodata & \nodata & 68.37\\

21 &\nodata &\nodata & \nodata &\nodata& 74.184 [1] & \nodata & \nodata & \nodata\\

22 &\nodata &\nodata & \nodata &\nodata& 80.602 [1] & \nodata & \nodata& \nodata\\

23 & \nodata  & \nodata & \nodata & \nodata&87.263 [1] & \nodata & \nodata & \nodata\\

24 & \nodata  & \nodata & \nodata & \nodata&94.259 [1] & \nodata & \nodata & \nodata\\

25 & \nodata & \nodata & \nodata & \nodata &101.25 [1]  & \nodata & \nodata & \nodata\\

26 & (55.410 [5]) & \nodata & 55.10 [6] & 56.01 & 108.64 [2] & \nodata &
106.09 [2] & 108.18\\
\enddata
\tablenotetext{a}{2DHF \citep{ivanov00}} 
\tablenotetext{b}{1DHF \citep{neuhauser86}}
\tablenotetext{c}{DF \citep{jones_pb85_2}}
\tablecomments{All energies are in keV. $\Delta \tilde
z=(2-5)\times 10^{-2}$. $N_{iter}=5$. Other atomic models computed several ground state energies. (1)
Thomas-Fermi-Dirac method \citep{skjervold84}: 4.14, 7.73 [keV] for
$Z=6$ at $B_{12}=1.0, 5.0$. 56.21, 105.89 [keV] for
$Z=26$ at $B_{12}=1.0, 5.0$. (2) Restricted variational method \citep{flowers77}: 0.545, 0.913
[keV] for $Z=2$ at $B_{12}=1.0, 5.0$. 53.13, 101.7 [keV] for
$Z=26$ at $B_{12}=1.0, 5.0$. (3) Density functional method with correction
\citep{koessl88} provide lower energy value 110.4 [keV] for $Z=26$ at $B_{12}=5.0$.}
\end{deluxetable}

\begin{deluxetable}{cccccc}
\tabletypesize{\small}
\tablewidth{0pt}
\tablecaption{Oscillator strengths of bound-bound transitions for hydrogen. \label{h_os}}
\tablehead{
\colhead{$(m,\nu) \rightarrow (m',\nu')$} & $\beta=5$  & $\beta=10$ &
$\beta=50$ & $\beta=100$ & $\beta=500$}
\startdata
(0,0) $\rightarrow$ (1,0) & 0.115 & 7.15\ttwo & 2.29\ttwo &  1.46\ttwo & 5.04\tthr\\
 & (0.108) & (6.88\ttwo) & (2.32\ttwo) & (1.44\ttwo) & (5.06\tthr)\\
\\
(0,0) $\rightarrow$ (0,1) & 0.627 & 0.553 & 0.437& 0.364& 0.228\\
 & (0.615) & (0.560) & (0.414) & (0.351) & (0.227)\\
\\
(0,0) $\rightarrow$ (0,3) & 5.28\ttwo & 4.93\ttwo& 4.19\ttwo& 3.64\ttwo& 2.47\tthr\\
 & (5.24\ttwo)&(4.94\ttwo) & (3.99\ttwo) & (3.52\ttwo) & (2.45\tthr)\\
\\
(1,0) $\rightarrow$ (2,0) & 8.80\ttwo&5.47\ttwo&1.97\ttwo&1.30\ttwo&5.47\tthr\\
 &(8.46\ttwo)&(5.47\ttwo)&(1.98\ttwo)&(1.30\ttwo)&(5.48\tthr)\\
\\
(0,1) $\rightarrow$ (0,2) & 1.53 & 1.50 & 1.38 & 1.33 & 1.20\\
 & (1.52) & (1.48) & (1.38) & (1.33) & (1.20)\\ 
\enddata
\tablecomments{Grid points are the same as in the tables of energy values. The values in parentheses are from multi-configurational
 Hartree-Fock method \citep{ruder}. For $\Delta m = 1$ transitions, finite
nuclear mass correction is included.}
\end{deluxetable}

\begin{deluxetable}{cccccc}
\tablewidth{0pt}
\tablecaption{Oscillator strengths of bound-bound transitions for
helium.\label{he_os}}
\tablehead{
\colhead{$(m,\nu)$} & $\beta_Z=5$ & $\beta_Z=10$ & $\beta_Z=50$ &
$\beta_Z=100$ & $\beta_Z=500$}
\startdata
(1,0) $\rightarrow$ (1,1) & 0.523 & 0.424 & 0.240 & 0.187 & 0.103\\
 & (0.500) & (0.410) & (0.239) & (0.184) & (0.099) \\
\\
(1,0) $\rightarrow$ (2,0) & 7.52\ttwo& 4.56\ttwo& 1.51\ttwo& 9.27\tthr&2.85\tthr\\
 & (7.35\ttwo) &  (4.64\ttwo)& (1.54\ttwo)&  (9.36\tthr) &  (2.85\tthr)\\
\\
(2,0) $\rightarrow$ (2,1) & 0.609 & 0.522 & 0.335 & 0.253 & 0.133\\
 & (0.598) & (0.511) & (0.318) & (0.250) & \nodata\\
\\
(2,0) $\rightarrow$ (3,0) & 4.45\ttwo & 2.86\ttwo &
9.69\tthr  & 6.00\tthr & 1.90\tthr\\
 & (4.50\ttwo) & (2.87\ttwo) & (9.74\tthr) &
(6.02\tthr) & (1.90\tthr)\\
\\
(3,0) $\rightarrow$ (3,1) & 0.663 & 0.582 & 0.380 & 0.308 & 0.145\\
 & (0.661) & (0.577) & (0.375) & (0.299) & \nodata\\
\enddata
\tablecomments{Grid points are the same as in the tables of energy values. The values in parentheses are from 1DHF \citep{ruder}.}
\end{deluxetable}

\begin{deluxetable}{cccc}
\tablewidth{0pt}
\tablecaption{Oscillator strengths of bound-bound transition $(m,0)
\rightarrow (m,1)$ for carbon. \label{c_os}}
\tablehead{
\colhead{$m$} & $\beta=200$ & $\beta=500$  & $\beta=1000$
}
\startdata
0 & 4.10\ttwo & 1.30\ttwo & 5.81\tthr \\
 & (3.97\ttwo) & (1.11\ttwo) & (5.71\tthr) \\
\\
1 & 5.98\ttwo & 2.25\ttwo & 1.28\ttwo\\
 & (5.67\ttwo) & (2.31\ttwo) & (1.35\ttwo) \\
\\
2 & 7.97\ttwo &  3.92\ttwo & 2.47\ttwo \\
 & (8.00\ttwo) & (3.80\ttwo) & (2.36\ttwo) \\
\\
3 & 0.114 & 6.11\ttwo & 3.77\ttwo \\
 & (0.109) & (5.65\ttwo)& (3.62\ttwo)\\
\\
4 & 0.156 & 8.63\ttwo & 5.72\ttwo\\
 & (0.144) &(7.94\ttwo) & (5.21\ttwo)\\
\\
5 & 0.197 & 0.116 & 7.83\ttwo\\
 & (0.193) & (0.113)& (7.55\ttwo)\\
\enddata
\tablecomments{The values in parentheses are from 1DHF \citep{miller90}. The
other electrons are in their ground state.}
\end{deluxetable}

\end{document}